\documentclass{aa}  
\usepackage{graphicx}
\usepackage{txfonts}
\usepackage{lipsum}
\usepackage{subcaption}
\usepackage{lscape}
\usepackage{placeins}

\usepackage[colorlinks=true, linkcolor=blue, citecolor=blue, urlcolor=blue, pdfstartview=FitH]{hyperref}

\makeatletter
\@ifundefined{switchlinenumbers}{}{
  
  \let\linenumbers\relax
}
\makeatother

\newcommand{\bef}{\begin{figure}}      
\newcommand{\eef}{\end{figure}}      
\newcommand{\bea}{\begin{eqnarray}}    
\newcommand{\eea}{\end{eqnarray}}      
\newcommand{\be}{\begin{equation}}      
\newcommand{\ee}{\end{equation}}

\newcommand{\RNum}[1]{\uppercase\expandafter{\romannumeral #1\relax}}

\usepackage{color}

\newcommand{\ve}[1]{\mathbf{q1}}

\newcommand{\av}[1]{\ensuremath{\left\langle q1 \right\rangle}}

\newcommand{\tve}[1]{\tilde{\boldsymbol{q1}}}

\def\bse{\begin{subequations}}
\def\ese{\end{subequations}}

\begin{document}
\nolinenumbers
\makeatletter
\makeatother

\titlerunning{Large-Scale Galaxy Correlations from the DESI DR1}
\authorrunning{Sylos Labini \& Antal}

   \title{Large-Scale Galaxy Correlations from the DESI First Data Release}

   \author{Francesco Sylos Labini
          \inst{1,2}\fnmsep\thanks{sylos@cref.it} and  Tibor Antal\inst{3} 
                 }

   \institute{Centro Ricerche Enrico Fermi,  I-00184, Roma, Italia
         \and
            INFN Unit\`{a} Roma 1, Dipartimento di Fisica, Universit\`{a} di  Roma Sapienza, I-00185 Roma, Italia 
            \and
            School of Mathematics and the Maxwell Institute for Mathematical Sciences, University of
            Edinburgh, Edinburgh, United Kingdom
             }

   \date{Received xxxx; accepted yyyy}

\abstract{
We quantify galaxy correlations using two distinct three-dimensional samples from the first data release of the \textit{Dark Energy Spectroscopic Instrument} (DESI): the Bright Galaxy Sample (BGS) and the Luminous Red Galaxy Sample (LRGS). Specifically, we measure the \emph{conditional average density}, defined as the average density of galaxies observed around a typical galaxy in the sample. To minimize boundary effects, we adopt a conservative criterion: only galaxies for which a spherical volume of radius $r$, centered on them, is fully contained within the survey footprint are included in the computation.
For the BGS, we construct four volume-limited subsamples in order to eliminate biases arising from luminosity-dependent selection effects. By contrast, the LRGS is approximately volume-limited by design. The resulting samples span different depths, providing an opportunity to test the stability of statistical measurements across survey volumes of increasing size.
Our results show that the conditional average density follows a power-law decay, $\langle n(r) \rangle \propto r^{-0.8}$, without exhibiting any transition to homogeneity within the survey volume. The large statistics of the DESI samples also allow us to demonstrate that finite-size effects become significant as $r$ approaches the boundaries of the sample volumes. Consistently, we find that the distribution of density fluctuations follows a Gumbel distribution—characteristic of extreme-value statistics—rather than a Gaussian distribution, which would be expected for a spatially homogeneous field.
These findings confirm and extend the trends previously observed in smaller redshift surveys, supporting the conclusion that the galaxy distribution does not undergo a transition to spatial homogeneity within the probed scales, up to $r \sim 400$~\text{Mpc}/$h$.
}
   \maketitle

\section{Introduction}

The spatial distribution of galaxies is a fundamental observable for probing the large-scale structure of the universe. Accurately characterizing its statistical properties poses a significant challenge, as this distribution exhibits complex, highly inhomogeneous features and includes the largest structures ever detected in the cosmos  representing one of the most compelling manifestations of complexity in nature (see, e.g., \cite{deLapparent_etal_1986,Geller+Huchra_1989,Gott_etal_2005,Tully_etal_2014,Valade_etal_2024}).

A key question, debated for decades \citep{deVaucouleurs_1970,Pietronero_1987,Lemson+Sanders_1991,Coleman+Pietronero_1992,Davis_1997,Pietronero_etal_1997,SylosLabini_etal_1998,Wu_etal_1999,Joyce_etal_2005,Hogg_etal_2005,SylosLabini_etal_2007,Sarkar_etal_2009,SylosLabini_2011,Scrimgeour_etal_2012,Pandey_2013,Alonso_etal_2014,Whitbourn+Shanks_2014,SylosLabini_etal_2014,Conde-Saavedra_etal_2015,Alonso_etal_2015,Sarkar+Pandey_2015,Pandey+Sarkar_2015,Shirokov_etal_2016,Laurent_etal_2016,Park_etal_2017,Ntelis_etal_2017,Goncalves_etal_2018b,Goncalves_etal_2018a,Heinesen_2020,Teles_etal_2021,DeMarzo_etal_2021,Kim_eta_2022,Andrade_etal_2022,Dias_etal_2023,Valade_etal_2024,Pashapour-Ahmadabadi_etal_2025,Einasto_2025b,Einasto_2025a}, concerns the scale at which the galaxy distribution can be considered statistically uniform. Determining this scale is essential for assessing whether the observed structures are compatible with current models of galaxy formation. Addressing this issue requires statistical methods that are free of strong assumptions and based on the most conservative hypotheses, in order to minimize luminosity selection biases and finite-size effects. In this respect, one must consider several subtle points. 

The first requirement is that the statistical measure should not rely on the assumption that the average density is well-defined within a given sample --- an assumption which, operationally, implies that the sample volume is sufficiently large so that the average density measured in any of its sufficiently large sub-volumes yields statistically consistent values, modulo small-amplitude fluctuations. For instance, if the survey volume is a sphere of radius $R$, the presence of a void or of a large-scale structure with a size comparable to a significant fraction of $R$ would cast doubt on the validity of this assumption. 
 In general, the presence of large-scale voids and/or extended structures necessitates a critical assessment of whether a well-defined average density exists within the sample.

This can be tested operationally by measuring the conditional average density, $\langle n(r) \rangle$, which quantifies the average density within a sphere of radius $r$ centered on a typical point of the distribution. Specifically, it is defined as the average of the local number density obtained by centering a sphere of radius $r$ on each point in the sample. This statistic does not assume \emph{a priori} the existence of a global mean density and therefore provides a robust characterization of spatial inhomogeneities and correlations in the most general case.

A second crucial requirement in the analysis of spatial correlations is that the sample under study must be free from significant  \textit{luminosity selection effects} and \textit{boundary condition effects}.
Luminosity selection effects arise, for example, when a sample is constructed by selecting all galaxies with apparent luminosity brighter than a given threshold. Such a selection introduces a bias, since at any fixed distance, intrinsically brighter galaxies are overrepresented compared to fainter ones. To mitigate this effect, one must construct \textit{volume-limited (VL) samples}, which include only galaxies that would be observable throughout the entire survey volume, thereby ensuring uniform selection in luminosity space (see, e.g., \cite{book} and references therein).

Boundary condition effects are related to the way the conditional density is computed. To adopt the most conservative approach, one must include as center-points only those galaxies for which a sphere of radius $r$ is fully embedded within the sample volume (see, e.g., \cite{book} and references therein).

Both types of selection effects impose stringent limitations on the effective number of galaxies that can be used for statistical analysis. Therefore, it is essential to work with large, three-dimensional samples that have uniform and contiguous angular coverage.

In analyses of galaxy distributions across various surveys using the conditional density method  it has been consistently found that, up to scales of approximately 20 Mpc/$h$ --- where statistical reliability is high due to the possibility of inscribing a sufficient number of independent spheres within the survey volume --- the {conditional average density of the} galaxy distribution follows a simple power-law behavior {$\langle n(r\rangle)\propto r^{-\gamma}$} with an exponent $\gamma \approx 1$. A spatially homogeneous distribution would correspond to $\gamma =0$.  At larger scales (up to $\sim$100 Mpc/$h$), the statistical evidence for the continuation of this power-law behavior becomes weaker, but there is no clear indication of a transition to spatial uniformity (see \cite{Joyce_etal_2005} and references therein). 

Notably, \cite{Hogg_etal_2005} analyzed the Luminous Red Galaxy (LRG) sample from the SDSS survey and found that a power-law with exponent $\gamma = 1$ provides a good fit to the conditional density up to at least 20~Mpc/$h$, covering approximately a decade in scale. At larger scales, the conditional density continues to decrease but with a shallower slope, flattening around $\sim 70$~Mpc/$h$ and remaining nearly constant up to the maximum probed scale of 100~Mpc/$h$. The transition between these regimes is gradual: the integrated conditional density at $\sim 20$~Mpc/$h$ remains about twice the asymptotic value measured at the largest scales. \cite{Hogg_etal_2005} interpreted this behavior as indicative of a slow crossover toward homogeneity occurring on scales $\gtrsim 100$~Mpc/$h$. However, this trend may also reflect the presence of long-range correlations without clear evidence for convergence to homogeneity within the observed volume \citep{Joyce_etal_2005}. As we will discuss below, a crucial issue in measuring the conditional density on scales comparable to the sample size is {to control for} finite-size effects, which can introduce significant systematic errors.

Subsequently, \cite{SylosLabini_etal_2007} investigated the average conditional galaxy density in the Bright Galaxy Sample of the Sloan Digital Sky Survey Data Release 7 (SDSS DR7). Consistent with earlier analyses by \cite{Joyce_etal_2005,Hogg_etal_2005}, they found that, over a broad range of scales, both the average conditional density and its variance exhibit non-trivial scaling behaviors. In particular, for scales $20 < r < 80$~Mpc$/h$, the conditional density displays a  weak  dependence on the system size. However, \cite{SylosLabini_etal_2007} identified non-trivial finite-size effects arising from the presence of large-scale structures that, at sufficiently large scales, hinder the determination of a truly representative volume average. Building on this, \cite{Antal_etal_2009} demonstrated that the density fluctuations are well described by a Gumbel distribution, characteristic of extreme-value statistics --- in clear contrast to the Gaussian distribution expected in a spatially homogeneous configuration. They also introduced tests to quantify finite-size effects on scales comparable to the survey volume, finding that such effects are particularly relevant in the case of the SDSS DR7 Bright Galaxy Sample.
Together, these results support the presence of long-range correlations and scale-invariant structures in the galaxy distribution, extending up to scales of at least $\sim 100$~Mpc$/h$.

Since then, many authors have mapped the large-scale structure of galaxies with the goal of placing constraints on the scale of homogeneity, employing a wide variety of observational datasets and statistical methodologies. However, results have been inconsistent: some studies report a relatively small homogeneity scale of $\lambda_0 \approx 50$-$70$~Mpc$/h$ (see, e.g., \cite{Sarkar_etal_2009,Scrimgeour_etal_2012,Ntelis_etal_2017,Dias_etal_2023}), while others find significantly larger values, such as $\lambda_0 \approx 140$~Mpc$/h$ \citep{Pandey+Sarkar_2015} or even $\lambda_0 \approx 250$~Mpc$/h$ \citep{Sarkar+Pandey_2015}. 
These discrepancies largely arise from the diversity of statistical techniques and data sets employed across the literature, as well as from the use of different definitions of the homogeneity scale itself. 

In particular, the methods used to investigate large-scale homogeneity include:

\begin{itemize}
\item Multifractal analysis of the galaxy distribution in VL subsamples from SDSS DR6 \citep{Sarkar_etal_2009};

{
\item Galaxy counts in magnitude-limited samples from deep, narrow pencil-beam surveys
at $z \approx 1$, with limited angular coverage — e.g. the WiggleZ Dark Energy Survey
\citep{Scrimgeour_etal_2012}. 
We note that the procedure adopted in the WiggleZ analysis,
which corrects counts-in-spheres in a non–volume-limited sample by normalizing to the
expectation for a homogeneous distribution with the same completeness, implicitly
assumes large-scale homogeneity. As a result, the method effectively builds in the
conclusion it seeks to demonstrate, amounting to a form of circular reasoning. 
This correction scheme has subsequently been adopted in later studies 
(e.g. \citealp{Goncalves_2021});
}

\item Reconstruction of the local density field from redshift distributions and number counts \citep{Whitbourn+Shanks_2014};

\item Statistical tests of large-scale homogeneity using photometric redshift surveys \citep{Alonso_etal_2014};

\item Entropy-based measures of inhomogeneity in SDSS DR12 to evaluate the scale of the transition to homogeneity \citep{Pandey_2013, Sarkar+Pandey_2015};

\item Fractal analysis of the galaxy distribution in the FORS Deep Field survey, although without systematic corrections for luminosity selection effects or boundary conditions \citep{Conde-Saavedra_etal_2015};

\item Use of the angular homogeneity index applied to the 2MASS Photometric Redshift Catalogue \citep{Alonso_etal_2015};

\item Homogeneity analysis in the redshift range $2.2 < z < 2.8$ using the BOSS quasar sample \citep{Laurent_etal_2016};

\item Counts-in-spheres analysis of the CMASS galaxy sample from the BOSS survey to probe the transition to homogeneity \citep{Ntelis_etal_2017};

\item Assessment of homogeneity using the SDSS-IV DR14 quasar sample over the redshift range $0.80 < z < 2.24$ \citep{Goncalves_etal_2018b}.

\item Investigation of scaling properties in galaxy super-cluster catalogs  through the Mandelbrot-Zipf law \citep{DeMarzo_etal_2021};

\item Measurement of the angular scale of cosmic homogeneity using the LRG sample from SDSS DR16 \citep{Andrade_etal_2022};

\item Estimation of the homogeneity scale using spectroscopic samples of blue galaxies from SDSS \citep{Dias_etal_2023};

\end{itemize}

In general, the differing conclusions reached in the literature can be attributed to the use of heterogeneous statistical tools and the absence of a unified treatment of key observational effects, such as luminosity selection biases, survey geometry, and boundary conditions. Crucially, many of these studies do not adopt the most conservative approach described above --- namely, the use of fully three-dimensional, VL samples, statistically unbiased estimators, and a rigorous control of both selection effects and boundary artifacts. As a result, the inferred values of the homogeneity scale in such analyses may be systematically biased by methodological limitations. This is particularly important in redshift surveys, which are inherently limited both in the number of objects and in the coverage of the redshift volume.

In this work, we analyze several samples from the first data release of the \textit{Dark Energy Spectroscopic Instrument} (DESI), which has surveyed an unprecedented cosmological volume. We focus on two specific data sets: the Bright Galaxy Sample (BGS) and the Luminous Red Galaxy Sample (LRGS). From the BGS, we construct a series of volume-limited (VL) subsamples containing galaxies of different absolute luminosities and with progressively increasing depths, whereas the LRGS is already approximately volume-limited and benefits from high-precision photometric calibration. The resulting set of samples spans a wide range of depths, providing an opportunity to test the stability of statistical measurements across survey volumes of increasing size.

DESI is a next-generation spectroscopic survey installed on the 4-meter Mayall Telescope at Kitt Peak National Observatory. It is equipped with 5,000 robotically controlled optical fibers and covers a wide spectral range (360-980~nm), enabling simultaneous observations of thousands of targets. DESI's primary scientific goal is to map the large-scale structure of the Universe by obtaining redshifts for tens of millions of galaxies, quasars, and Lyman-$\alpha$ forest systems over approximately 14,000~deg$^2$. 
The first-year data release (DR1) of DESI includes over five million high-quality redshifts spanning the range $0.1 < z < 2.1$. Owing to its unprecedented combination of survey volume, redshift coverage, and spectroscopic data quality, DESI currently represents the most powerful galaxy clustering survey to date \citep{Hahn_etal_2023,Ross_etal_2025,Adame_etal_2025b,Adame_etal_2025a,DESICollaboration_2025}. 

However, to date, no studies have investigated the \textit{conditional density} in DESI data. Instead, analyses of clustering have primarily relied on the standard two-point correlation function \citep{Adame_etal_2025b,Adame_etal_2025a}, a statistical tool that presupposes that large-scale homogeneity of the galaxy distribution is reached in the samples under consideration \citep{book}. Exploring alternative, less assumption-dependent statistics --- such as the conditional density --- may thus provide complementary insights into the nature of cosmic structure and the validity of the homogeneity hypothesis.

The paper is organized as follows. In Sect.~\ref{sec:methods}, we detail the statistical methods used to measure the relevant quantities in the survey. In Sect.~\ref{sec:samples}, we describe the construction of the samples. Sect.~\ref{sec:results} presents the results of the analysis of the conditional density and its probability distribution function. Finally, in Sect.~\ref{sec:concl}, we discuss our findings in the context of standard cosmological models and summarize our main conclusions.

\section{Methods}
\label{sec:methods} 

The standard definition of the two-point correlation function,
\begin{equation}
\label{eq:xi}
\xi(r) = \langle \delta (\mathbf{r}) \, \delta (\mathbf{0}) \rangle \;, 
\end{equation}
relies on the assumption that the average density $n_0$ is well-defined and strictly positive in the infinite-volume limit. Here, the density contrast is defined as
\begin{equation}
\label{eq:df}
\delta (\mathbf{r}) = \frac{n(\mathbf{r}) - n_0}{n_0} \,.
\end{equation}
However, for systems exhibiting long-range correlations, it is possible that $n_0 = 0$ in the infinite-volume limit, which renders the definition of $\xi(r)$ ill-defined and physically meaningless.

In any finite sample of volume $V$ containing $N$ particles, the sample average density $n_s = N/V$ is necessarily positive. The crucial question is whether $n_s$ remains stable as $V$ increases ---that is, whether the system reaches statistical homogeneity within the sample volume. This issue can only be rigorously tested using unnormalized statistical quantities, such as the conditional average density $\langle n(r) \rangle$, which do not rely on the a priori assumption of a nonzero $n_0$ (see, e.g., \cite{book}, and references therein).

\subsection{Conditional galaxy density}

To address the issue of an inhomogeneous galaxy distribution, we will condition the local densities to be measured around galaxies. Let us denote the number of galaxies in a ball of radius $r$ around galaxy $i$ as $N_i(r)$, where $i=1,\dots, C(r)$, and $C(r)$ is the number of galaxies which can act as a center of a ball. Then the conditional galaxy density around galaxy $i$ is
\[
 n_i(r) = \frac{3N_i(r)}{4\pi r^3} \;.
\]
Note that this density is typically different for each center $i$, and it has an empirical distribution of densities.

The simplest quantities characterizing this density distribution is its mean
\begin{equation}
\label{eq:avecon}
 \langle n(r) \rangle = \frac{1}{C(r)} \sum_{i=1}^{C(r)} n_i(r)
\end{equation}
and its variance
\begin{equation}
\label{eq:variance} 
\Sigma^2(r) = \frac{1}{C(r)} \sum_{i=1}^{C(r)} \left(n_i(r) - \langle n(r) \rangle\right)^2 \,.
\end{equation}
For an uncorrelated homogeneous point distribution, like a Poisson point process with $N(r)$ number of galaxies in a ball of radius $r$, one has $\langle N(r) \rangle = \mathrm{Var}[ N(r) ]\propto r^3$, which leads to 
$\langle n(r) \rangle \approx \mathrm{constant}$ and
$\Sigma^2(r) \propto r^{-3}$, 
moreover the fluctuations become Gaussian for large $r$.

For a correlated but still homogeneous point distribution, the conditional density remains 
$\langle n(r) \rangle \approx \mathrm{const.}$, 
and one focuses on the  behavior of the normalized variance 
\[
\sigma^2 =  \frac{\langle n^2 (r) \rangle - \langle n(r) \rangle^2}{\langle n(r) \rangle^2 } = \frac{\Sigma^2(r)}{\langle n(r) \rangle^2} \;. 
\]
This 
depends on the nature of the correlations between density fluctuations (see Eq.\ref{eq:xi}). 
For instance, if the power spectrum of density fluctuations 
$P(k)$  --- i.e. the Fourier conjugate of $\xi(r)$ ---behaves for $k \rightarrow 0$ as 
$P(k) \propto k^{\,n}$ (with $n > -3$ to ensure integrability), then one finds \citep{glasslike,book}
\[
\sigma^2(r) \propto
\begin{cases}
r^{-(3+n)}  & \text{if } -3 < n < 1 \\[4pt]
r^{-4}\,\log r & \text{if } n = 1 \\[4pt]
r^{-4}  & \text{if } n > 1 \;\; ,
\end{cases}
\]
where $n=0$ corresponds to the uncorrelated (Poisson) case.  
The case $n=1$ corresponds to the large-scale correlations characteristic of standard cosmological models, for which the normalized mass variance decays faster than in an uncorrelated density field.  
In addition, for $-3 < n < 0$ the correlation function behaves as 
$\xi(r) \propto r^{-(3+n)}$ at large scales, and in this regime the normalized mass variance decays more slowly than for an uncorrelated density field.

Note that 
the average conditional density 
at scales smaller than the average nearest–neighbor distance 
is approximately related to the behavior of the nearest–neighbor probability density $\omega(r)$,  
i.e.\ the probability $\omega(r) dr$ that the nearest neighbor of a given galaxy lies at a distance in the interval $[r, r+dr]$, through the relation 
\begin{equation}
\label{eq:nn_relation}
\langle n(r) \rangle = \frac{\omega(r)}{4\pi r^2 \left(1- \int_0^r \omega(s)\,ds \right) }\; .
\end{equation}
Indeed, for the nearest galaxy being in $[r, r+dr]$, there should be no galaxy closer, but there should be one in the shell $[r, r+dr]$, and by assuming independence of these two events we get
$$
 \omega(r) dr \approx \left(1-\int_0^r \omega(s)\,ds\right) \times 4\pi r^2 \langle n(r) \rangle dr
$$
which leads to Eq.\ref{eq:nn_relation}.

For a Poisson point process with rate with $\langle n(r) \rangle = n_0$, the probability of not finding a galaxy in a ball of radius $r$ is just $e^{-n_0 4\pi r^3/3}$, and by differentiation we get
\begin{equation}
\omega(r) = 4\pi r^2 n_0 \exp\!\left( -\frac{4\pi n_0 r^3}{3} \right) \; .
\end{equation}
which indeed satisfies Eq.\ref{eq:nn_relation}. 
In contrast, for a correlated distribution such that 
$\langle n(r) \rangle \propto r^{-\gamma}$,  
the short–distance tail behaves as 
\begin{equation}
\omega(r) \propto r^{2-\gamma} \; ,
\end{equation}
which satisfies Eq.\ref{eq:nn_relation} for $\int_0^r \omega(s)\,ds\ll 1$.

When $\langle n(r) \rangle \approx \mathrm{const.}$, 
the probability density function (PDF) of the local density $n_i(r)$ is close to Gaussian,  
and thus one focuses on the PDF of the density fluctuations $\delta(r)$ (see Eq.~\ref{eq:df})  
to characterize small deviations. 

In contrast, when $\langle n(r) \rangle \propto r^{-\gamma}$  
the PDF of fluctuations is not expected to be Gaussian, and one expects that large deviations occur with higher probability than in a Gaussian process. 

As mentioned above, one of the key challenges in measuring the conditional density in a galaxy survey is the proper treatment of boundary conditions. It is important to note that $C(r)$ in Eqs.\ref{eq:avecon}-\ref{eq:variance} depends on the scale $r$ itself. A galaxy can be used as a center only if the sphere of radius $r$ centered on it is fully contained within the survey volume. If this condition is not satisfied, the corresponding count is excluded from the average, as the probed volume would not be spherical, thus introducing potential systematic distortions.

Indeed, when the sampling volume is not spherical, it may exhibit significantly different extensions along different directions, making it nontrivial to define the effective scale $r$ that enters into Eq.~\ref{eq:avecon}. This geometric asymmetry complicates the interpretation of the measured density, and may bias the estimate of $\langle n(r) \rangle$ if not properly accounted for. For this reason, a strict boundary condition is applied: only spheres entirely embedded within the survey volume are considered in the computation of the conditional density.

As $r$ approaches the radius of the largest sphere that can be entirely embedded within the survey geometry, two effects arise: (i) the number of usable centers $C(r)$ decreases rapidly, and (ii) the remaining centers tend to cluster near the geometric center of the sample. In this regime, although there may still be multiple centers contributing to the average, they all probe nearly the same spatial region. As a result, the estimate of the average density becomes dominated by a small and spatially localized part of the sample, leading to potential systematic effects due to the loss of spatial representativity  (see \cite{SylosLabini_etal_2014} and references therein). 

In addition to the number of usable centers, $C(r)$, we monitor in our analysis the quantity denoted as $f_{\mathrm{overlap}}$, defined as the ratio between the total overlapping volume of all intersecting sphere pairs and the total volume of all spheres:
\begin{equation}
\label{eq:overlap}
f_{\mathrm{overlap}} =
\frac{V_{\mathrm{overlap, total}}}{V_{\mathrm{sphere, total}}} \, .
\end{equation}
This parameter does not represent the fraction of the total spatial volume actually covered by the spheres, but rather quantifies the degree of redundancy arising from mutual overlaps.
In practice, $f_{\mathrm{overlap}} = 0$ indicates that all spheres are disjoint, whereas $f_{\mathrm{overlap}} = 1$ corresponds to a configuration where the total overlapping volume equals the sum of all sphere volumes—meaning that each unit of volume is, on average, doubly covered.
When $f_{\mathrm{overlap}} \gg 1$, the same spatial regions are intersected multiple times by different spheres.
In summary, $f_{\mathrm{overlap}}$ measures the degree of pairwise volume redundancy rather than the effective filling fraction, and values much larger than unity correspond to highly overlapping configurations.
From a statistical standpoint, a large value of $f_{\mathrm{overlap}}$ implies a significant correlation among neighboring measurements, thereby reducing their effective independence. This limits the reliability of estimators—such as averages or variances—derived from such overlapping volumes. In particular, if structures or voids exist on scales comparable to the size of the sample, a high degree of overlap may introduce a systematic effect that dominates the measurement, rather than merely contributing to its statistical uncertainty.

Finally, the {\it minimal scale} at which the determination of the conditional density and its moments remains statistically meaningful is set by a fraction  the average distance between nearest neighbors $\Lambda$. This characteristic scale, together with the full nearest-neighbor distance distribution, will be carefully computed and analyzed to assess the statistical validity of the measurements.

It should be noted that, in the analysis presented below, the statistical errors on the conditional density are computed as standard errors on the mean. However, at values of $r$ comparable to the size of the sample volume, the dominant source of uncertainty is systematic, rather than statistical. Therefore, it is essential to perform finite-size tests by varying the sample volume and carefully examining the stability of the results. This procedure allows us to assess the robustness of the observed trends and to identify possible biases arising from the survey’s geometry or selection effects.

In the following subsections, we will analyze not only the average and the variance, but also the full PDF of $n_i(r)$, in order to obtain a more complete statistical characterization of the galaxy distribution.

\subsection{The Gumbel distribution} 

\noindent

As this will be useful for the analysis that follows, we briefly recall the main properties of the Gumbel distribution, which is one of the three classical \textit{extreme–value distributions} (see, e.g., \citealt{Antal_etal_2009} and references therein).
The Gumbel distribution describes the probability of the maximum (or minimum) among a set of independent random variables drawn from a parent distribution whose PDF decays faster than any power law, such as an exponential distribution.
It therefore naturally arises in systems where extreme fluctuations are governed by exponentially bounded statistics.

\noindent
The density (PDF) of the Gumbel distribution, at a given radius $r$, is given by 
\begin{equation}
\label{eq:gumbel}
P(n) = \frac{1}{\beta} 
\exp\!\left[-\left(\frac{n - \mu}{\beta}\right) - \exp\!\left(-\frac{n - \mu}{\beta}\right)\right],
\end{equation}
where $n=n(r)$, $\mu$ is the location parameter and $\beta > 0$ is the scale parameter.

\noindent
By introducing the normalized variable
\begin{equation}
x = \frac{n - \mu}{\beta},
\end{equation}
the PDF takes the universal form
\begin{equation}
f(x) = \exp\!\left[-x - e^{-x}\right].
\end{equation}
{with} cumulative distribution $e^{-e^{-x}}$.
The mean and variance of the Gumbel distribution \eqref{eq:gumbel} are
\begin{align}
\langle n \rangle &= \mu + \gamma \beta, \\[4pt]
\Sigma^2 &= \frac{\pi^2 \beta^2}{6},
\end{align}
where $\gamma \simeq 0.5772$ is the Euler-Mascheroni constant.

The Gumbel distribution thus provides a simple yet powerful statistical model to describe fluctuations governed by extreme-value statistics, in contrast to the Gaussian distribution expected for homogeneous random fields.

\section{The samples}
\label{sec:samples}

The DESI Data Release 1 includes several spectroscopic galaxy samples targeting different redshift ranges: the Bright Galaxy Sample (BGS) at low redshifts, $0.1 < z < 0.4$; the Luminous Red Galaxy Sample (LRGS) in the intermediate redshift range, $0.4 < z < 1.1$; the Emission Line Galaxy Sample (ELGS) covering $0.8 < z < 1.6$; and the Quasar Sample (QSOS), spanning $0.8 < z < 2.1$ \citep{Adame_etal_2025a}. In the present analysis, we focus on the first two samples (BGS and LRGS), leaving the investigation of the ELGS and QSOS samples for forthcoming work.

As mentioned above, a key requirement for the computation of the conditional density is the selection of a sample volume with a regular geometry. Given the nature of the observational data, this geometry is best represented by a spherical sector, i.e., a portion of a sphere whose radial depth corresponds to the maximum extent of the sample, and whose angular coverage matches the largest contiguous region of the sky covered by the survey.

\subsection{Sample geometry and boundary conditions}

\begin{figure} 
\includegraphics[width=0.45\textwidth]{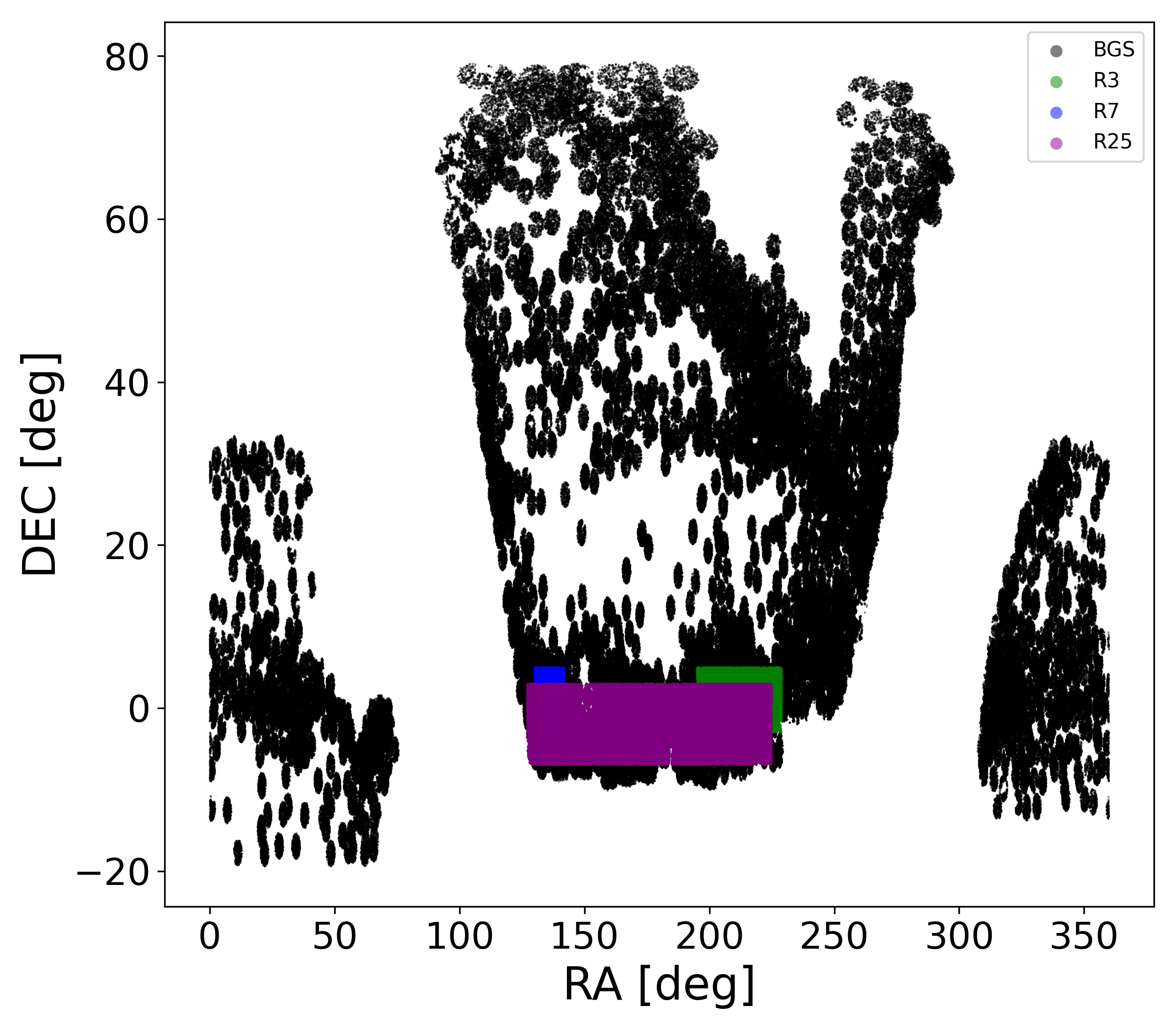}
\includegraphics[width=0.45\textwidth]{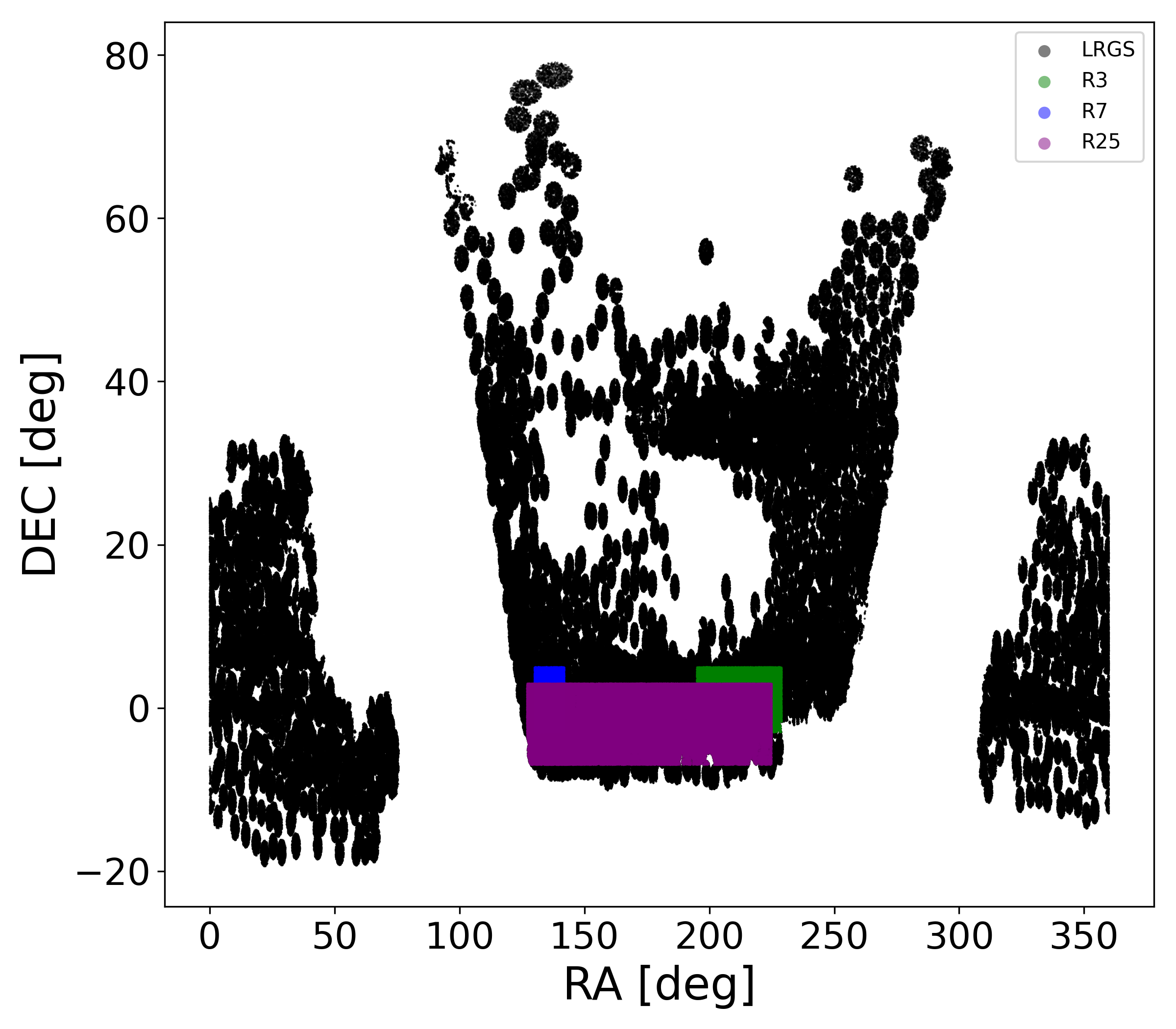}\\
\caption{DESI DR1 angular projection of the DESI~DR1 Bright Galaxy Sample (BGS) and Luminous Red Galaxy Sample (LRGS), together with the selected angular regions used in this analysis.
}
\label{angproj}
\end{figure}

The angular mask of the DESI~DR1 sample is highly irregular, reflecting the complex geometry of the survey footprint.
To construct a subsample suitable for measuring the conditional density, we identify several contiguous regions within the survey where the angular coverage can be considered  approximately uniform.
Among these, region~R25 is the largest, while regions~R3 and~R7 correspond to sub parts of R25 that also extend slightly beyond its boundaries.
The angular limits defining all selected regions are reported in Table~\ref{table:ang}.
\begin{table}
\caption{Limits Right Ascension and Declination (in degrees) and completeness of the three angular regions selected.}
\label{table:ang} 
\begin{centering} 
\begin{tabular}{|c | c | c | c | c | c |} 
\hline 
Region  & $\alpha_1$ & $\alpha_2$ & $\delta_1$ & $\delta_2$ & compl \\ 
\hline 
R3 & 195  &   229 &    -3  &    5 & 0.985 \\   
R7 &   130   &  142 &    -3   &   5  & 0.959 \\   
R25 & 127   &  225   &  -7   &   3   & 0.958\\   
\hline 
\end{tabular}
\end{centering} 
\end{table}
Figure~\ref{angproj} shows the full angular projection of the DESI~DR1 BGS and LRGS, together with the selected angular regions used in this analysis.

To estimate the completeness of each angular region, we proceed as follows.  
The probability that the DESI instrument successfully measures a redshift for a given target depends on its intrinsic properties, sky position, and redshift. This probability distribution is commonly referred to as the \textit{selection function}. It is typically anisotropic and exhibits a complex dependence on redshift.

To model and correct for these effects, we use the DESI catalogs that associate observed galaxies with corresponding \textit{random catalogs}, which Poisson-sample the same survey volume.
These random catalogs are weighted and subsampled to reflect the selection effects present in the data. In doing so, they account for the survey footprint, spatial variations in completeness (e.g., due to the number of observations per field), instrumental performance, and other observational systematics.

To evaluate the completeness of each angular region, we analyze the DESI random catalogs, which incorporate both the angular survey mask and the redshift-dependent selection function defined over the observed footprint. Specifically, we divide the angular footprint into $N_{\mathrm{RA}}$ bins in right ascension and $N_{\sin(\delta)}$ bins in $\sin(\delta)$, where $\delta$ is the declination. This binning produces:
\[
N_{\mathrm{tot}} = N_{\mathrm{RA}} \times N_{\sin(\delta)}
\]
angular cells.

Each cell is bounded in right ascension by $(\alpha_1, \alpha_2)$ and in declination by $(\sin\delta_1, \sin\delta_2)$. With this scheme, all cells subtend the same solid angle:
\begin{equation}
\label{omega}
\Omega_{\mathrm{cell}} = (\alpha_2 - \alpha_1) \times (\sin\delta_2 - \sin\delta_1) = \Delta_{\mathrm{RA}} \times \Delta_{\sin(\delta)} \,.
\end{equation}

Due to the irregular geometry of the DESI angular mask, not all cells are equally covered. For each well-sampled cell $i$, we compute the local density of random points as:
\[
n_i = \frac{N_i}{\Omega_{\mathrm{cell}}} \,,
\]
where $N_i$ is the number of random points in cell $i$. Variations in $n_i$ reflect differences in angular completeness. 

We define the maximum observed density as:
\[
n^{\mathrm{max}} = \max_i \{ n_i \} \,.
\]
Then, the completeness of angular region $i$, with generic solid angle $\Omega_i \gg \Omega_{\mathrm{cell}}$, is estimated as:
\begin{equation}
\label{eq:compl} 
\mathrm{compl}_i = \frac{N_i}{\Omega_i} \cdot \frac{1}{n^{\mathrm{max}}} \,.
\end{equation}

In Table~\ref{table:ang}, we report the estimated completeness values for each of the selected angular regions. 
To determine $n^{\mathrm{max}}$, we applied a very fine  partition of the survey area, dividing it into 144 angular cells of equal solid angle. 
The maximum local random density, computed from this fine-grained grid, is then used in Eq.~\ref{eq:compl} to evaluate the completeness of the three larger angular regions employed in this study.

Note that, in order to ensure that selection effects are negligible within the selected angular regions, we have repeated the analysis of the conditional density, variance, and the PDF of fluctuations using the corresponding random catalogs. We found that, as expected, the conditional density and variance in the random samples are consistent with those of a uniform distribution. However, the PDF of fluctuations, while approaching a Gaussian shape at sufficiently small scales, may still deviate at scales comparable with the sample's sizes slightly due to residual effects of the survey selection function encoded in the construction of the random catalogs themselves.

\begin{figure*} 
\includegraphics[width=0.45\textwidth]{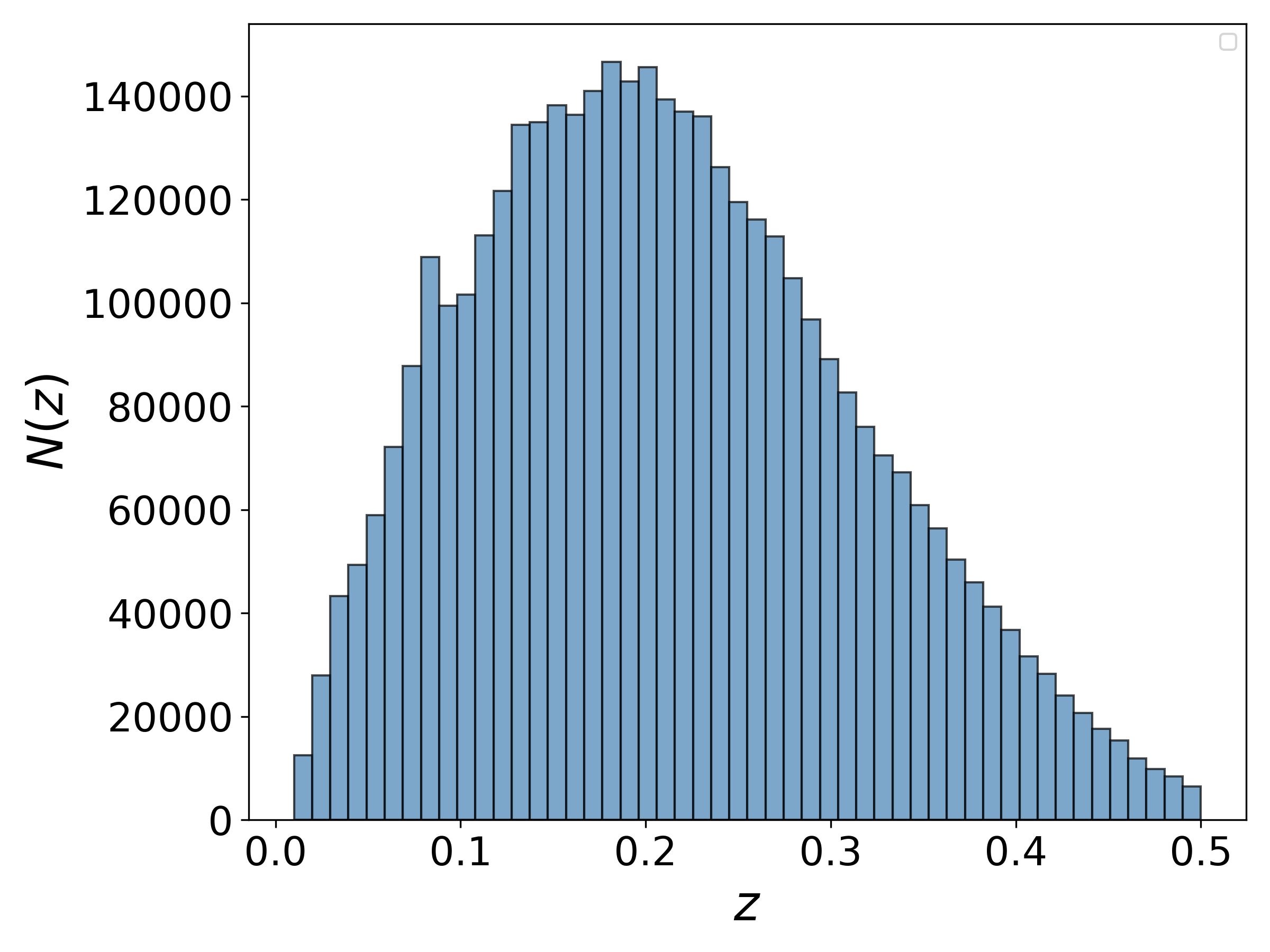}
\includegraphics[width=0.45\textwidth]{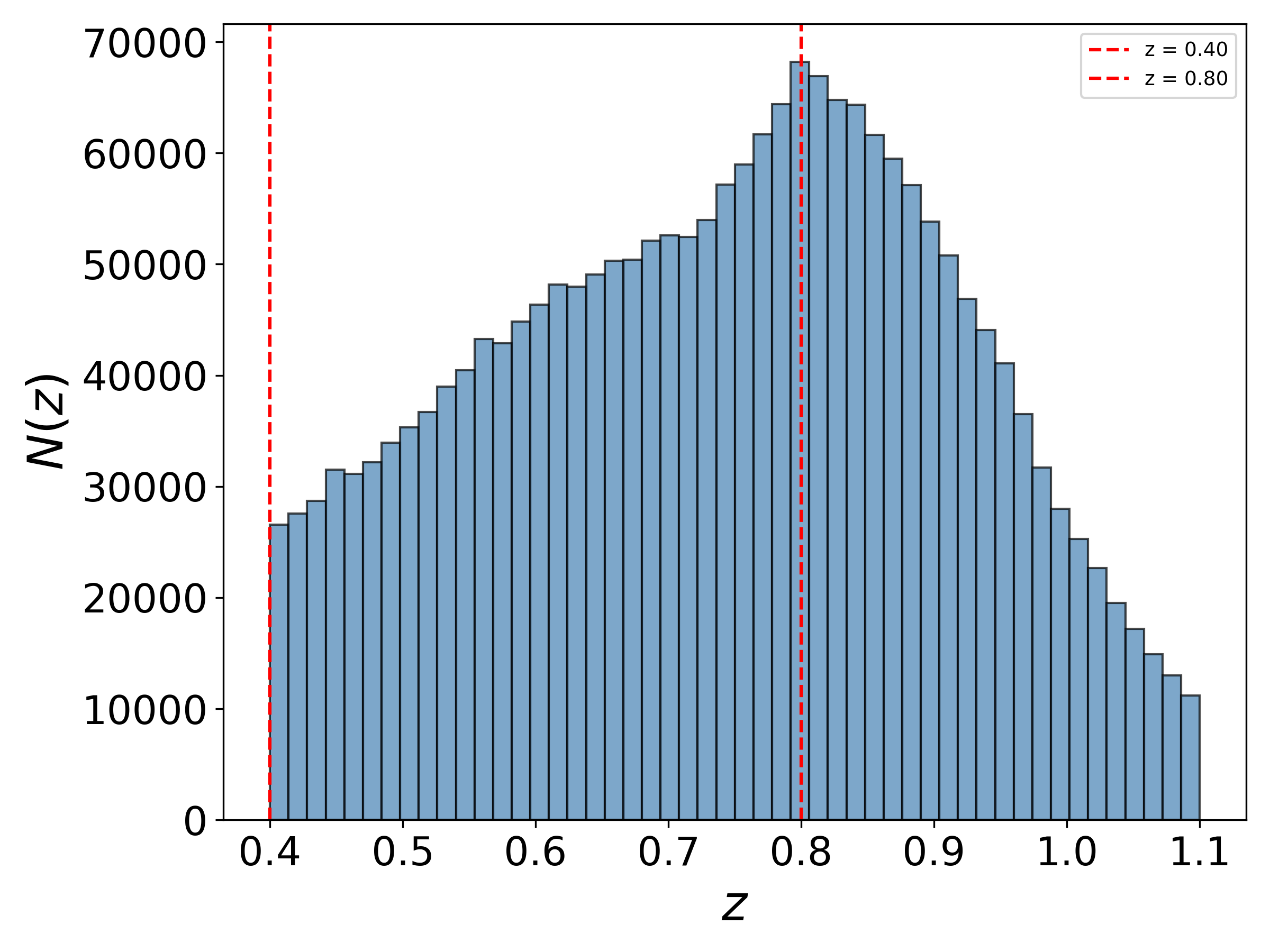}\\
\includegraphics[width=0.45\textwidth]{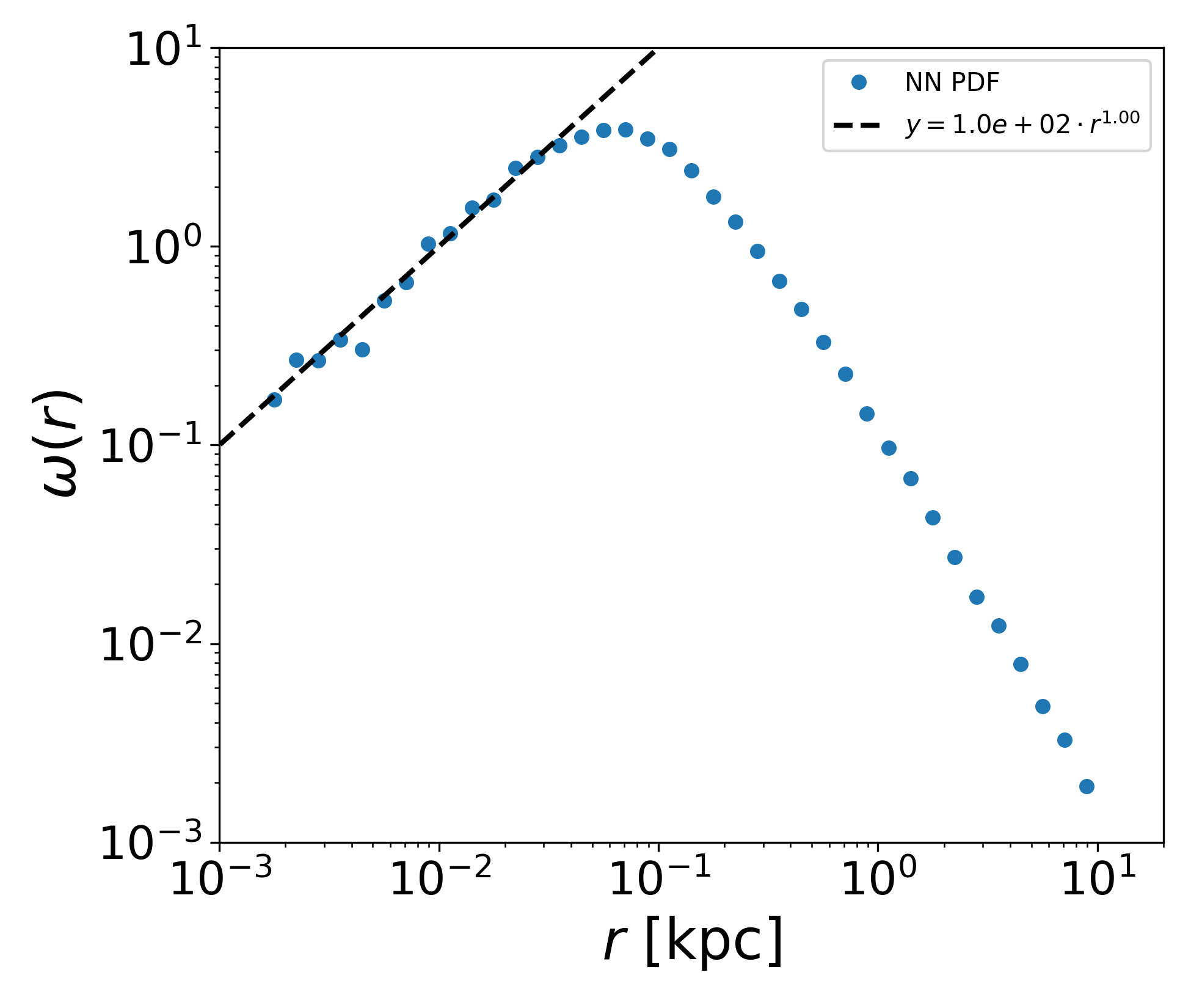}
\includegraphics[width=0.45\textwidth]{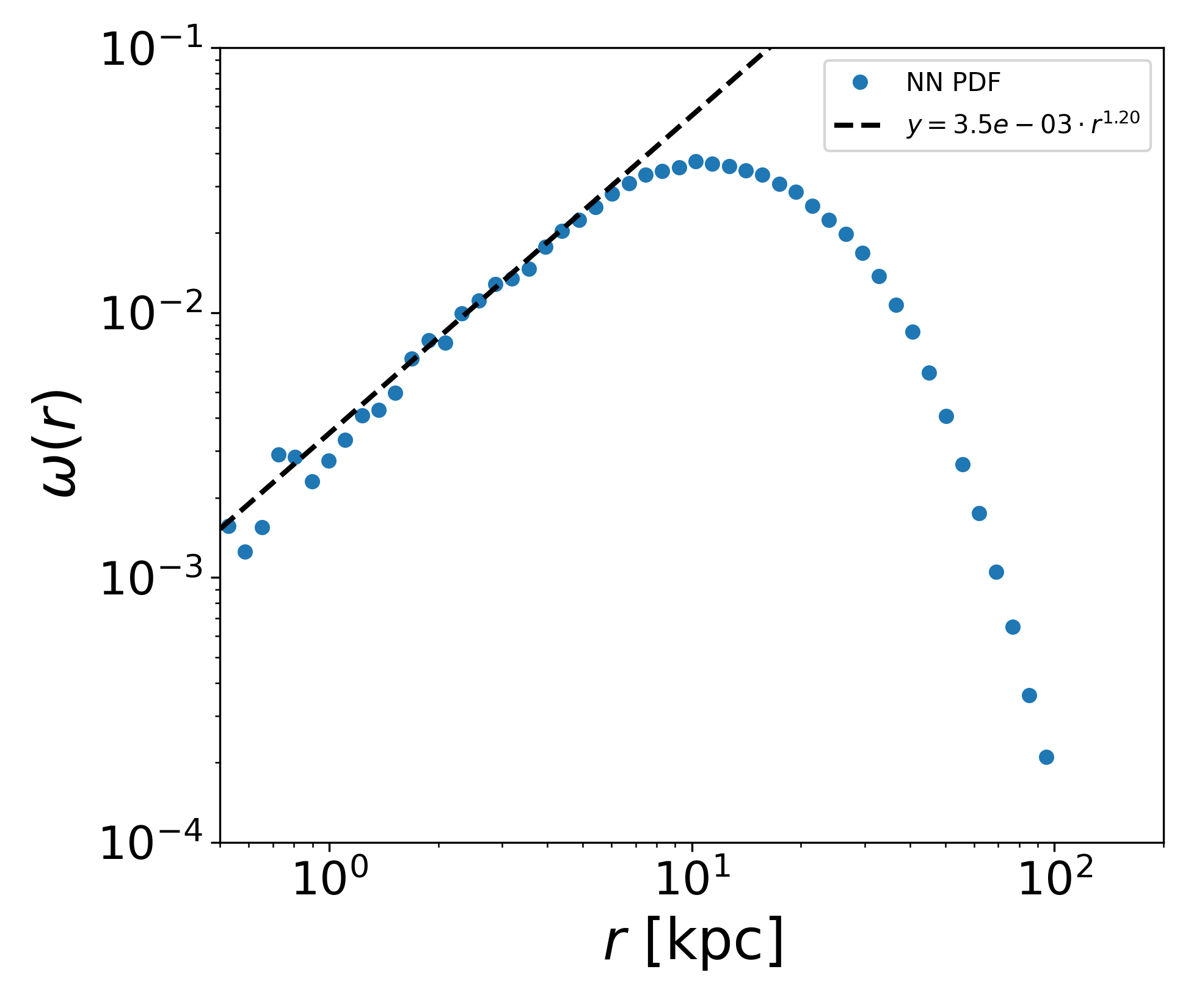}
\caption{ Upper panels: 
Redshift distribution of the BGS sample (left panel) and of the LRGS sample (right panel).
Bottom panels: Nearest neighbors distribution of the BGS sample (left panel) and of the LRGS sample (right panel).
} 
\label{pdf_bgs}
\end{figure*}

\begin{figure*} 
\includegraphics[width=0.45\textwidth]{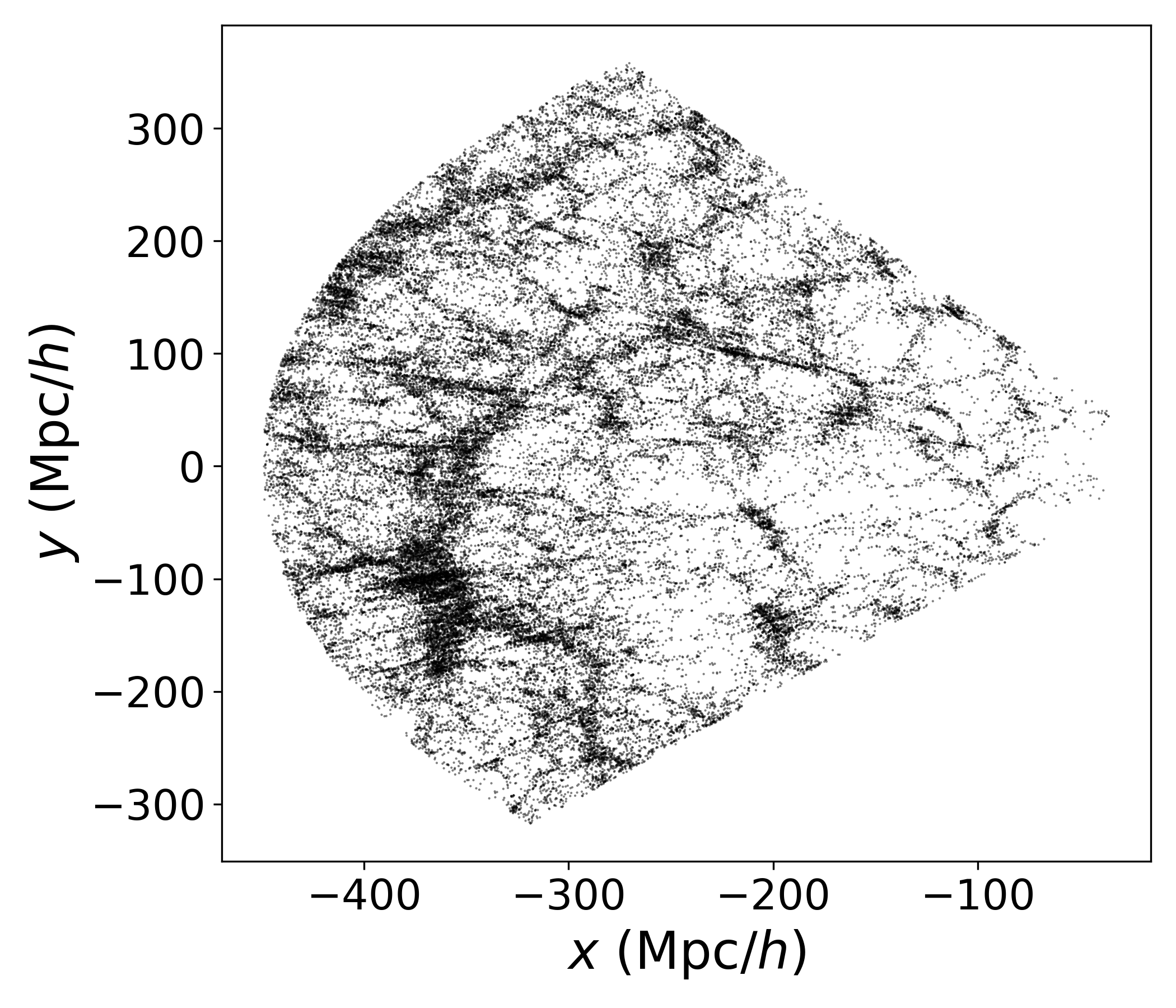}
\includegraphics[width=0.45\textwidth]{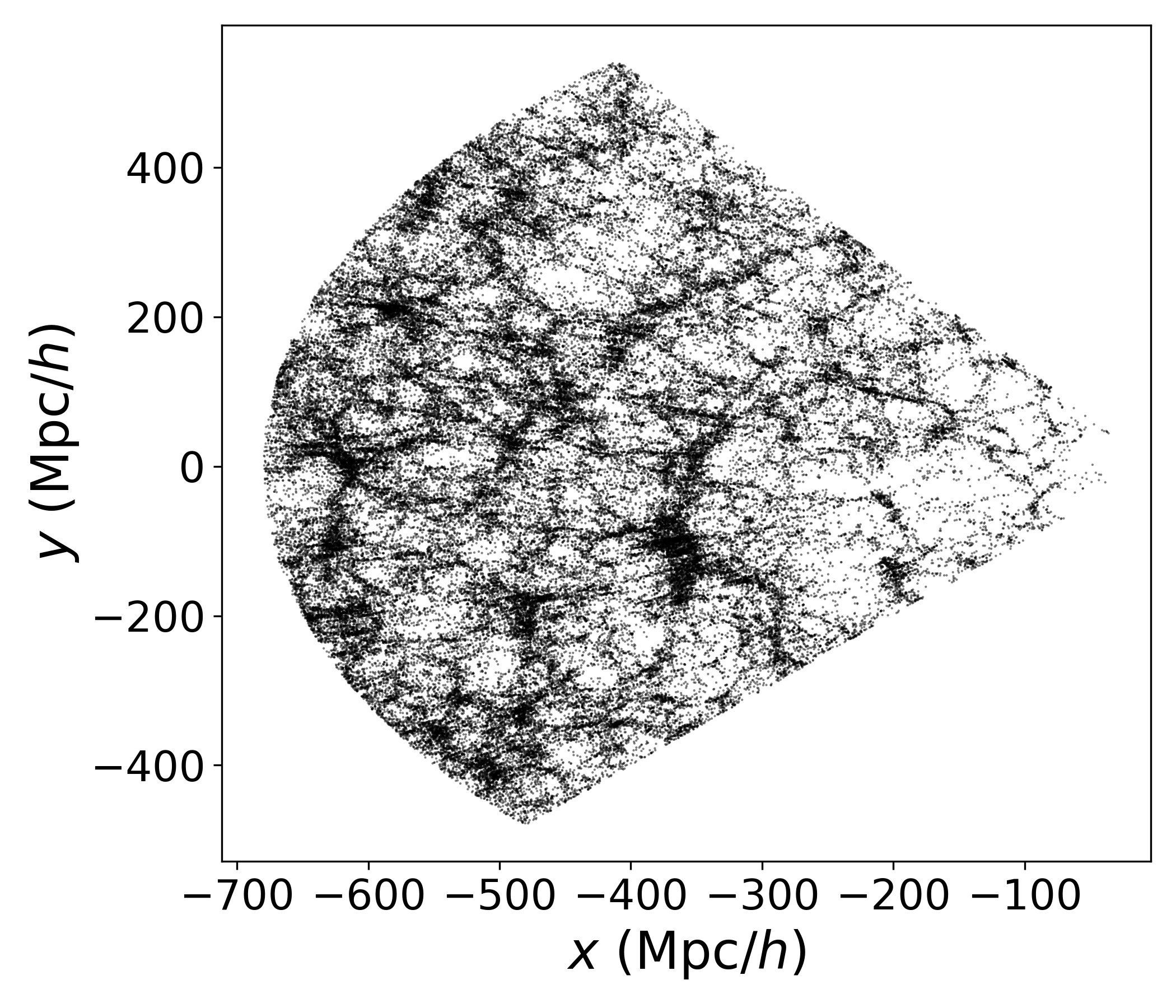}\\
\includegraphics[width=0.45\textwidth]{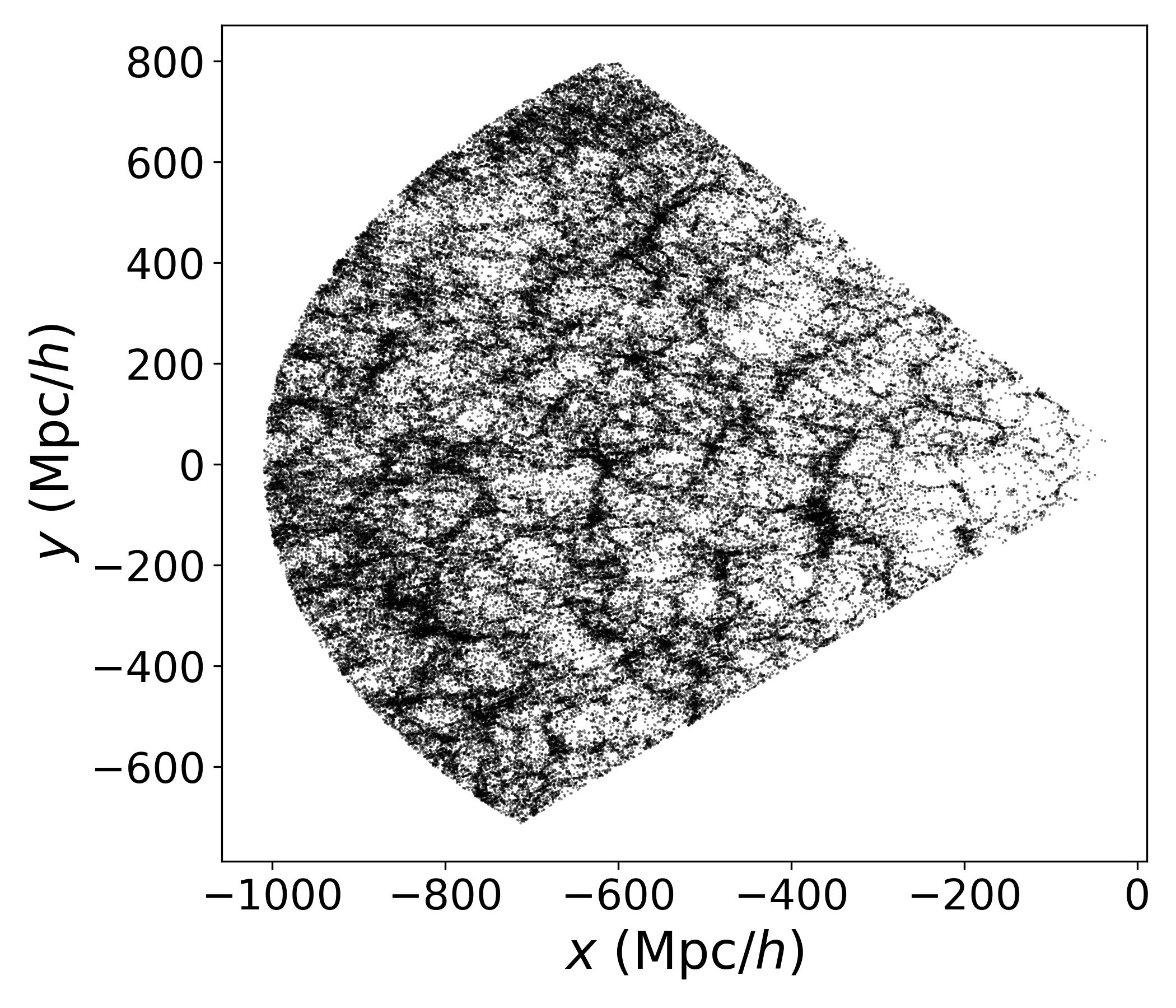}
\includegraphics[width=0.45\textwidth]{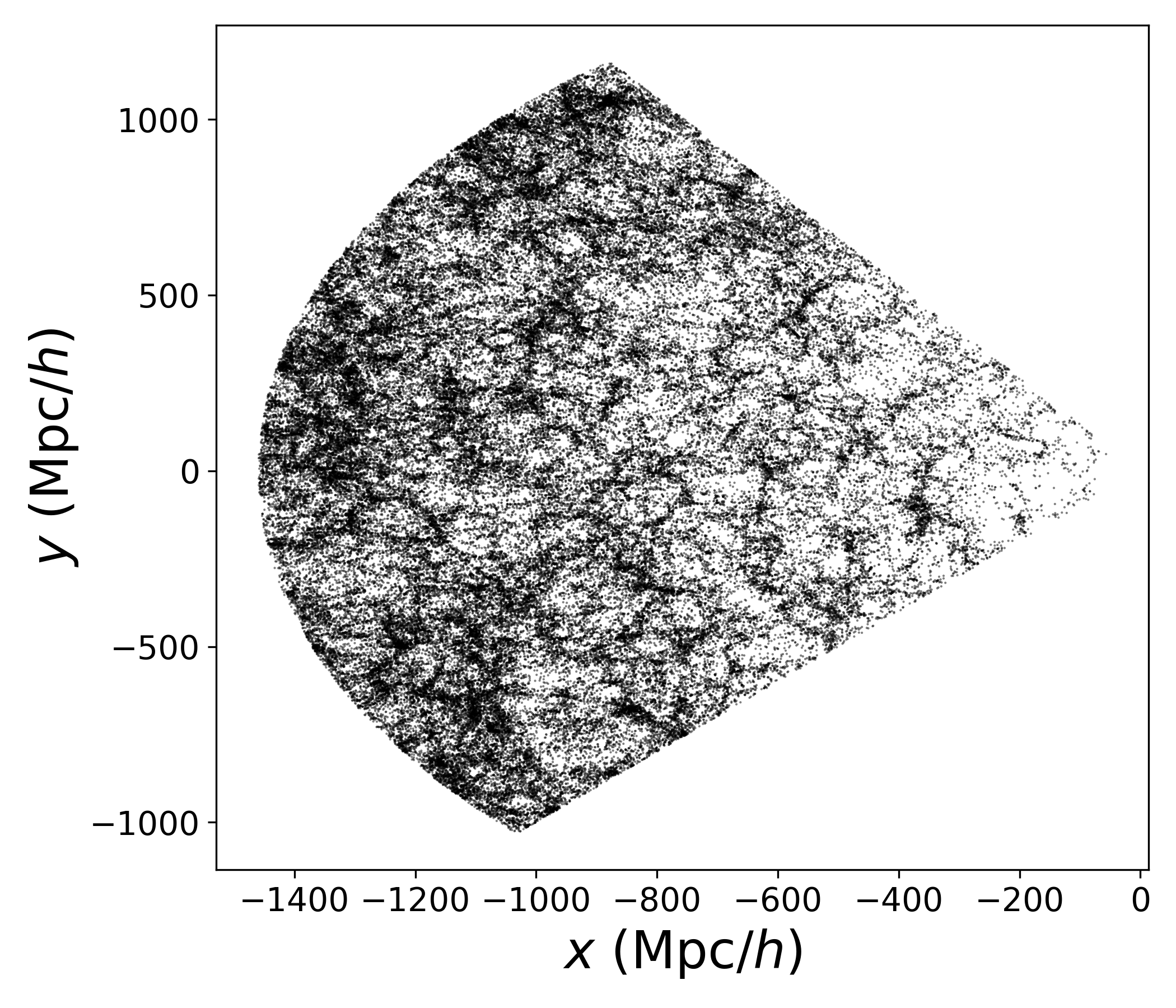}
\caption{Projection in Galactic coordinates on the $x$-$y$ plane of the four volume-limited samples in the region R25: VL2, VL3, VL4, and VL5, from left to right.} 
\label{xy_bgs}
\end{figure*}

\subsection{Bright Galaxy  Sample}

Figure~\ref{pdf_bgs} (top-left panel) shows the redshift distribution of the magnitude-limited BGS samples. The effects of luminosity selection are evident in the shape of the histograms: for a uniform distribution in a Euclidean universe, one would expect $N(z) \propto z^2$. Instead, the observed distribution peaks and declines, a feature determined by the survey's selection function.

To mitigate such luminosity-dependent selection effects, it is necessary to construct VL samples. These are defined such that, within a given range of absolute magnitudes and redshifts, the sample is complete ---  i.e., it includes all galaxies that would be observable above the survey's flux limit throughout the entire volume. Volume-limited samples are thus essential for performing unbiased statistical analyses of the spatial distribution of galaxies.

We first compute the apparent $r$-band magnitude, which is not directly provided in the clustering catalog but can be derived from the dereddened fluxes. 
The DESI catalogs provide fluxes in nanomaggies (e.g., \texttt{flux\_r\_dered}), and the conversion to magnitude is given by:
\[
m_r = 22.5 - 2.5 \log_{10}(\text{flux\_r\_dered}) \;.
\]
Assuming a standard flat $\Lambda$CDM cosmology with parameters 
$H_0 = 70 \, \mathrm{km\,s^{-1}\,Mpc^{-1}}$, $\Omega_m = 0.3$, and $\Omega_\Lambda = 0.7$, 
the luminosity distance is computed as:
\begin{equation}
d_L(z) = (1 + z) \, \frac{c}{H_0} \int_0^{z} 
\frac{dz'}{\sqrt{\Omega_m (1+z')^3 + \Omega_\Lambda}} \,,
\label{eq:dl_lcdm}
\end{equation}
where $c$ is the speed of light. 
From the luminosity distance, we calculate the distance modulus:
\[
\mu = 5 \log_{10}\left(\frac{d_L}{\text{pc}}\right) - 5 \,,
\]
and finally the absolute magnitude in the $r$-band:
\[
M_r = m_r - \mu \;.
\]
This procedure allows us to assign an absolute magnitude to each galaxy in the catalog, 
which is necessary for constructing VL samples and analyzing the luminosity distribution.

The $k$-corrected absolute magnitude in the $r$-band is given by:
\begin{equation}
M_r = m_r - 5 \log_{10} \left( \frac{d_L}{10\,\text{pc}} \right) - K_r(z) \,,
\end{equation}
where $K_r(z)$ is the $k$-correction to the rest-frame $r$-band.  
To compute $K_r(z)$, one can use the code originally developed by \cite{Blanton+Roweis_2007}, or rely on approximate methods based on observed colors and redshift.
A practical alternative is to use an empirical approximation, valid for low-redshift galaxies ($z \lesssim 0.3$), such as:
\begin{equation}
K_r(z, g-r) \approx 2.5 \cdot z - 1.5 \cdot (g-r) \cdot z \,,
\end{equation}
which has been shown to work reasonably well for SDSS BGS-like samples.

 Tab.\ref{table:vl} shows the properties of the VL samples and Tab.\ref{table:vl2}  the properties of the main angular regions we considered for the BGS. We also report the values  of the  average nearest-neighbor separation.
Figure~\ref{xy_bgs} shows the projection onto the $x$–$y$ plane of the R25 region for the different VL samples. As the volume increases ---  that is, moving from VL2 to VL5 --- large-scale structures become less visually prominent, although they remain discernible whereas 
  in Fig.\ref{pdf_bgs} (bottom-left panel) 
 it is shown the whole nearest-neighbor PDF.

 Note that our analysis has been performed in redshift space, i.e., we do not apply any correction to reconstruct real-space positions. However, these corrections are relevant only on small scales, i.e., for $r < 10$ Mpc$/h$, whereas the samples considered here cover a much larger range of scales.

\begin{table}
\caption{ Properties of the volume-limited samples: the first column lists the VL sample name; $D_{\mathrm{max}}$ is the maximum comoving distance (in Mpc/$h$); $M_r$ is the limiting absolute magnitude; $N$ is the total number of galaxies in the sample; and $\Lambda$ (in Mpc/$h$) denotes the average nearest-neighbor separation.}
\label{table:vl} 
\begin{centering} 
\begin{tabular}{|c | c | c | c| c| } 
\hline 
Sample & $D_{max}$ & $M_r$ & $N$ & $\Lambda$ \\ 
\hline 
VL2 & 450 & -19 & 361017&       1.2              \\  
VL3 & 680 & -20 & 647289&        0.7             \\  
VL4 & 1010 & -21 & 927335&       0.5              \\  
VL5 & 1460 & -22 & 845586&        0.4             \\  
\hline 
\end{tabular}
\end{centering} 
\end{table}

\begin{table}
\caption{ Number of galaxies $N$ in the different regions for the different volume-limited samples. }
\label{table:vl2}
\begin{centering} 
\begin{tabular}{|c | c | c | } 
\hline 
Sample &   Region   & $N$\\ 
\hline 
VL2&   R3    & 17489 \\  
VL2&   R7    & 4103 \\  
VL2 &  R25  & 58582 \\  
\hline 
VL3&   R3    & 31454 \\  
VL3&   R7    & 7222 \\  
VL3 &  R25  & 100209 \\  
\hline 
VL4&   R3    & 40724 \\  
VL4 &   R7    & 13457 \\  
VL4 &  R25  & 133960 \\  
\hline 
VL5 &   R3    & 38364 \\  
VL5 &   R7    & 13785 \\  
VL5 &  R25  & 125315 \\  
\hline 
\end{tabular}
\end{centering} 
\end{table}

\begin{figure*} 
\includegraphics[width=0.33\textwidth]{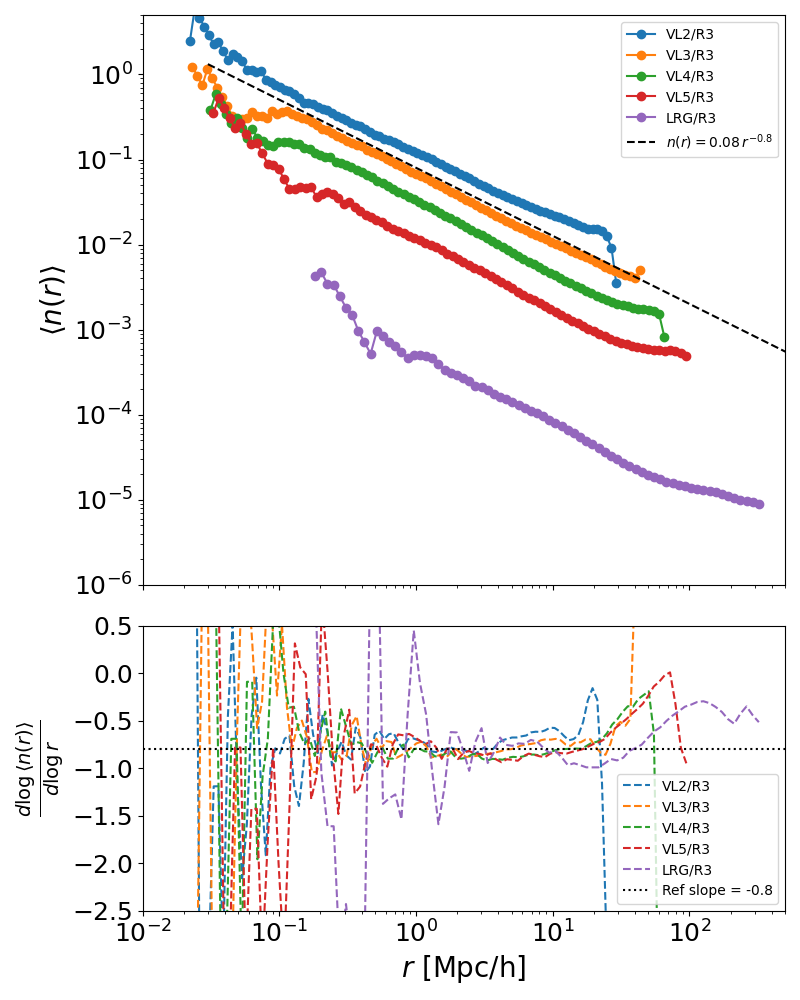}
\includegraphics[width=0.33\textwidth]{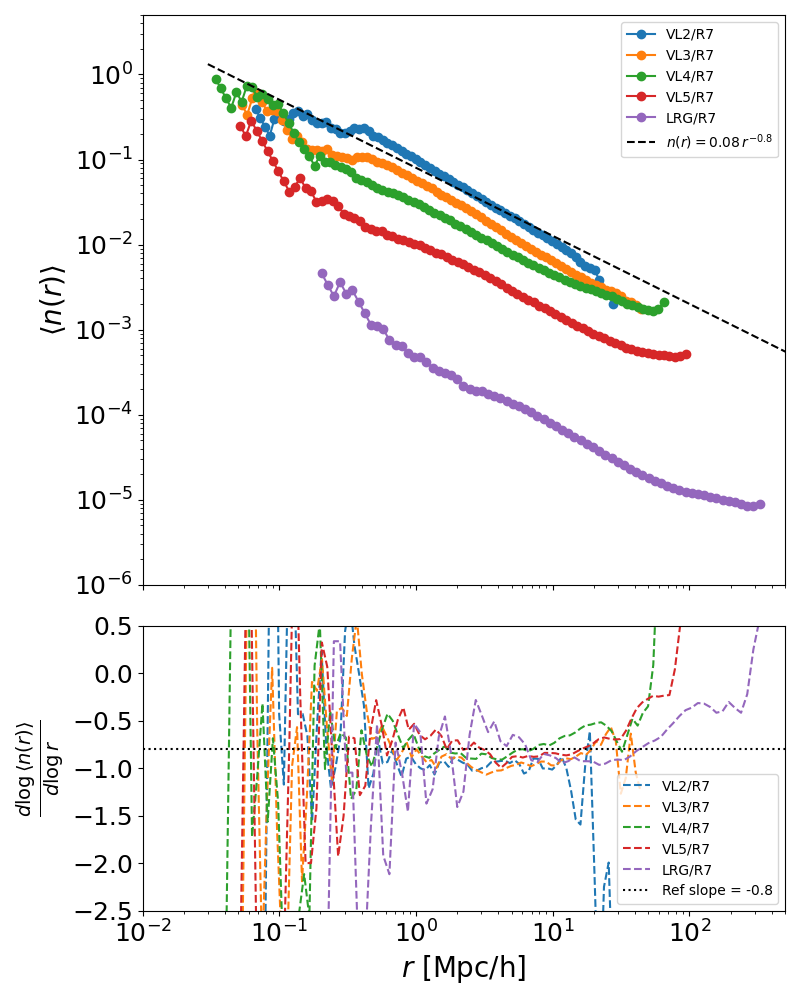}
\includegraphics[width=0.33\textwidth]{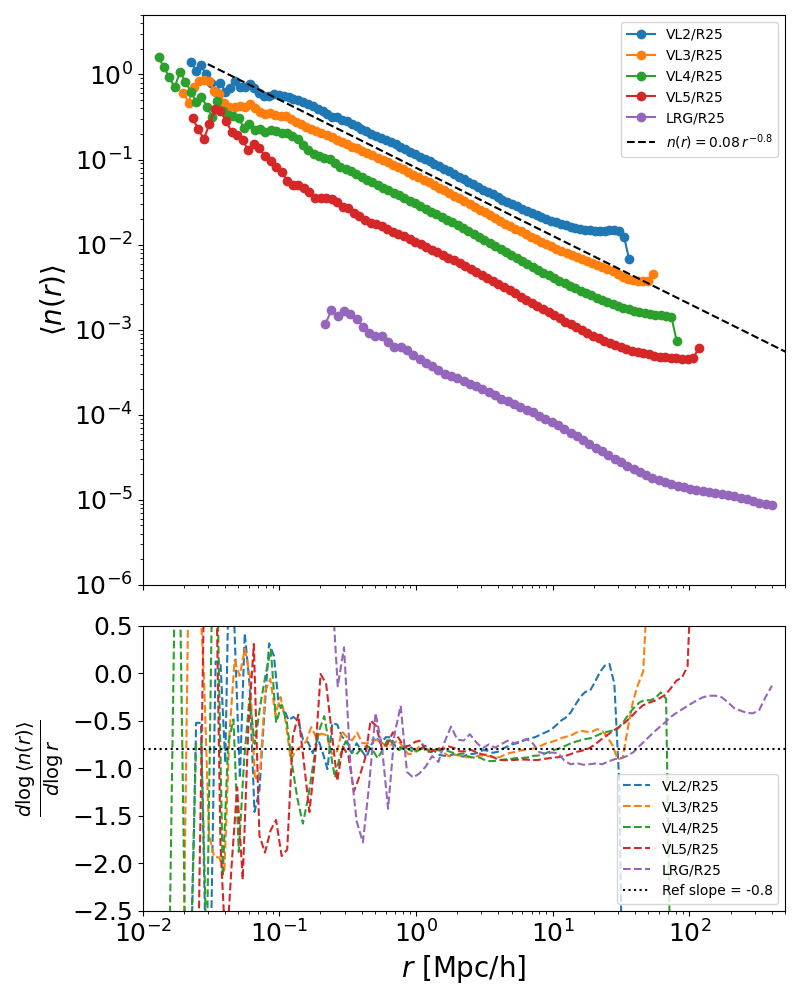}
\caption{Conditional average density $\langle n(r) \rangle$ (Eq.~\ref{eq:avecon}) measured in the four volume-limited subsamples --- VL2, VL3, VL4, and VL5 --- extracted from the BGS, as well as from the LRGS. Each panel corresponds to one of the three angular regions: R3, R7, and R25. In the bottom panels it is shown the logarithmic derivative.
} 
\label{SL_bgs_1}
\end{figure*}

\begin{figure} 
\includegraphics[width=0.5\textwidth]{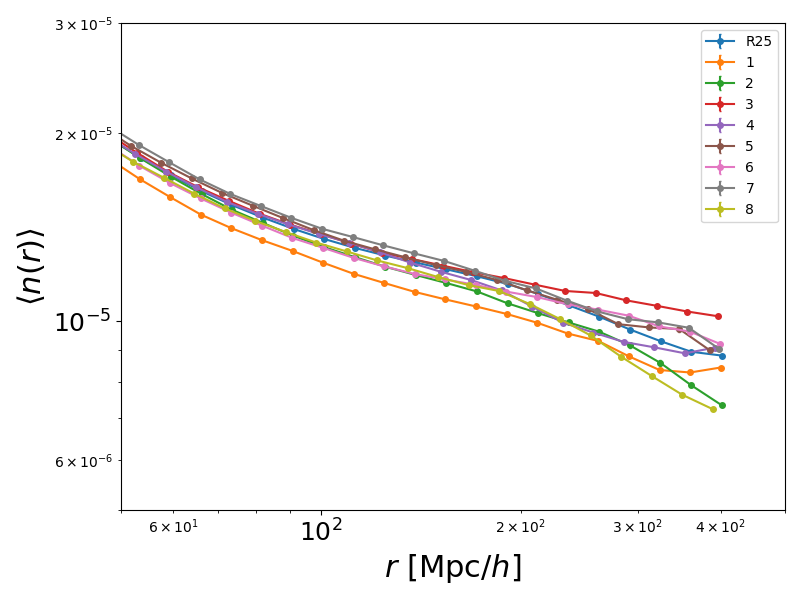}
\caption{Average conditional density for the LRGS 
in eight non-overlapping sub-regions of R25. One may note that non-negligible fluctuations, of order $\sim 20\%$, characterize the variation of the average conditional density among these sub-regions.
}
\label{SL_lrgs_2}
\end{figure}
\par

\subsection{Luminous Red Galaxy Sample}

The LRGS is approximately volume-limited; thus, in principle, no additional cuts in absolute magnitude or distance are required. The top-right panel of Fig.~\ref{pdf_bgs} shows the redshift distribution $N(z)$ of the LRGS. The distribution peaks at $z \approx 0.8$
and it extends up to $z \approx 1.1$ with a sharp cutoff at $z \approx 0.4$ ($d_L \approx 2100$~Mpc$/h$). 

The decline of $n(z)$ beyond $z \approx 0.8$ is primarily due to selection effects, which reduce the completeness of the sample at higher redshifts. For this reason, we define the redshift limits of our analysis as $0.4 < z < 0.8$ roughtly corresponding to a well-defined VL subsample. We have performed several tests by varying the redshift boundaries of the LRGS sample, and the results remain reasonably stable across these variations.

Tab.\ref{table:lrg} shows the properties of the LRGS in the same three  angular regions we considered for the BGS (see the bottom-right panel of Fig.\ref{pdf_bgs}  
for the whole nearest-neighbor PDF).  
\begin{table}
\caption{ Number of galaxies and average distance between nearest neighbors (in Mpc/$h$) in the three different angular regions for the Luminous Red Galaxy Sample.
}
\label{table:lrg} 
\begin{centering} 
\begin{tabular}{|c | c | c | } 
\hline 
Sample &   $N$  & $\Lambda$ [Mpc/$h$]\\ 
\hline 
R3 &  61218 & 21.8 \\  
R7 & 19723& 22.7 \\  
R25 &  206392 & 22.3 \\  
\hline 
\end{tabular}
\end{centering} 
\end{table}

\section{Results}
\label{sec:results}

\subsection{Average conditional density}

Figure~\ref{SL_bgs_1} displays the conditional average density $\langle n(r) \rangle$ (Eq.~\ref{eq:avecon}) measured in the four VL subsamples --- VL2, VL3, VL4, and VL5 --- extracted from the BGS, as well as one extracted from the  LRGS. Each panel corresponds to one of the three angular regions: R3, R7, and R25.
Each curve corresponds to a different subsample, defined by progressively increasing maximum depth and stricter absolute magnitude cuts. As the depth increases and the galaxies become intrinsically brighter, their spatial number density decreases and so the amplitude of $\langle n(r) \rangle$.

The average nearest--neighbor distance is $\Lambda \approx 1$~Mpc$/h$ for the BGS and $\approx 20$~Mpc$/h$ for the LRGS.  
The nearest--neighbor distribution (Fig.~\ref{pdf_bgs}) is fully consistent with the behavior of the average conditional density (Fig.~\ref{SL_bgs_1}): for $r \lesssim \Lambda$, the slope $\gamma \approx 0.8$ recovered from the conditional density agrees with the expectation derived from Eq.~\ref{eq:nn_relation}.

The comparison among these subsamples allows us to assess the robustness of the statistical signal against variations in sample depth and galaxy luminosity, as well as to identify the scales beyond which finite-size effects or selection biases may begin to influence the measurement. The observed consistency among the curves at intermediate scales,
i.e. $100$ Mpc/$h$ $\gtrsim r \gtrsim$ 1 Mpc/$h$,
confirms the stability of the conditional density in this regime, while deviations at large scales are primarily attributable to differences in sample volumes and boundary conditions.

We note that the amplitude of fluctuations in the conditional density decreases as the sample volume increases --- that is, when moving from R7 to R3 and to R25. This behavior reflects the improved statistical stability afforded by larger samples, where a greater number of independent centers and a wider spatial coverage lead to a smoother estimation of $\langle n(r) \rangle$.
\begin{figure} 
\includegraphics[width=0.5\textwidth]{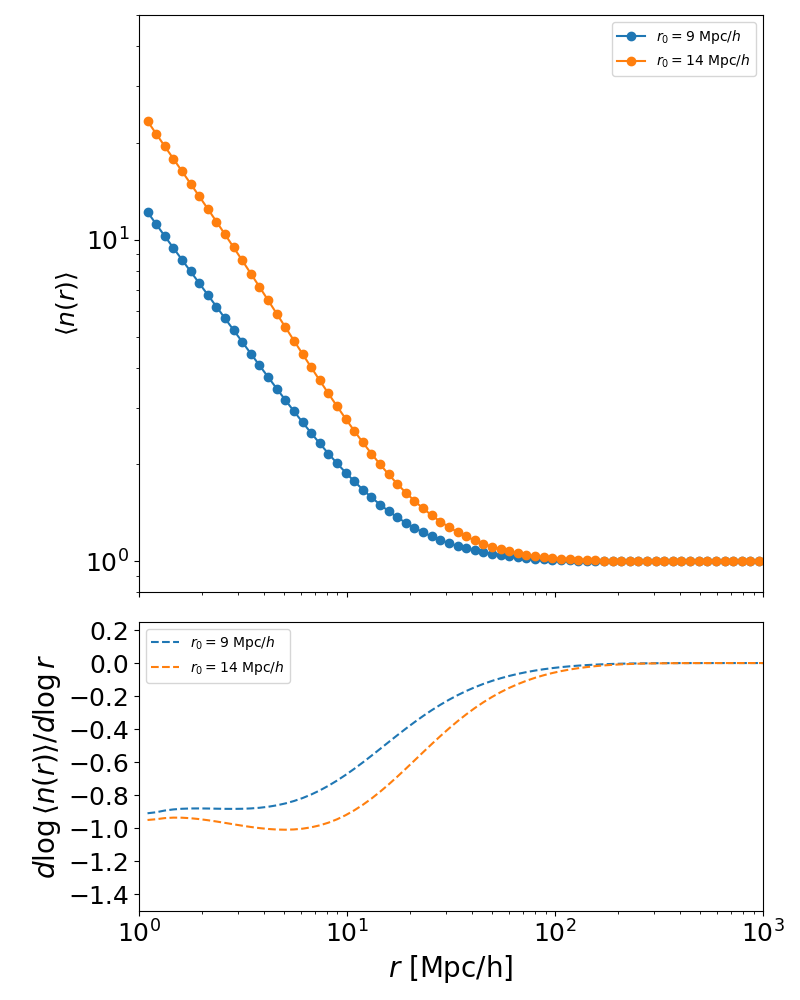}
\caption{Upper panel: behavior of the average conditional density (in units of the mean density $n_0$) in a CDM model  for different amplitudes, as characterized by the length scale $r_0$ defined by 
$\overline{\xi}(r_0) = 1$. Bottom panel: logarithmic slope. 
}
\label{SL_cdm}
\end{figure}
\par

\begin{figure} 
\includegraphics[width=0.5\textwidth]{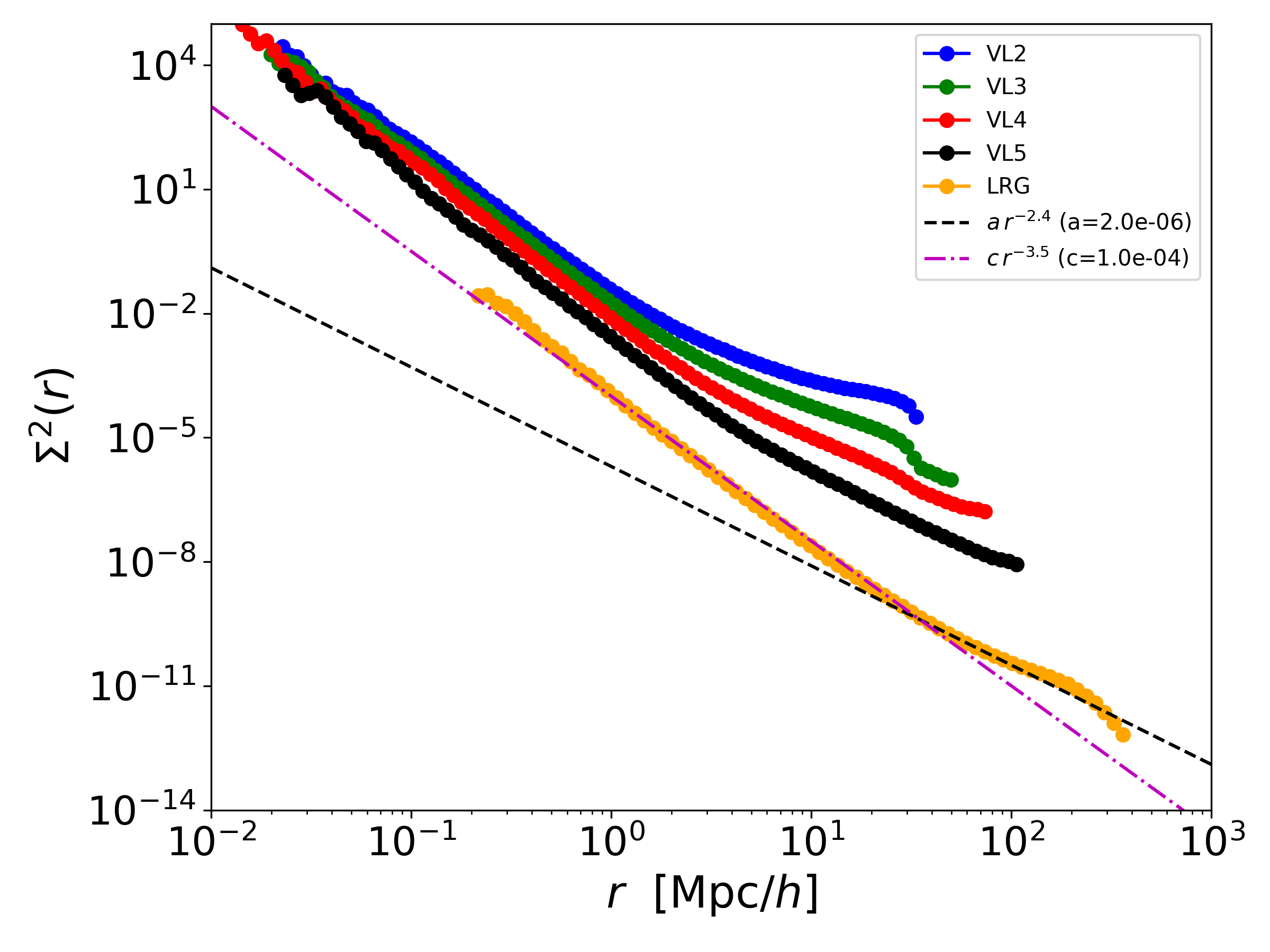}
\caption{
Variance $\Sigma^2(r)$ measured in the region R25 for the four VL subsamples (VL2, VL3, VL4, and VL5) extracted from the BGS, and for the LRGS sample (R25). Two reference lines 
proportional to $r^{-3.5}$ and $r^{-2.4}$ are shown.}
\label{SL_bgs_5}
\end{figure}
\par

\begin{figure}[t]
\centering
\includegraphics[width=0.5\textwidth]{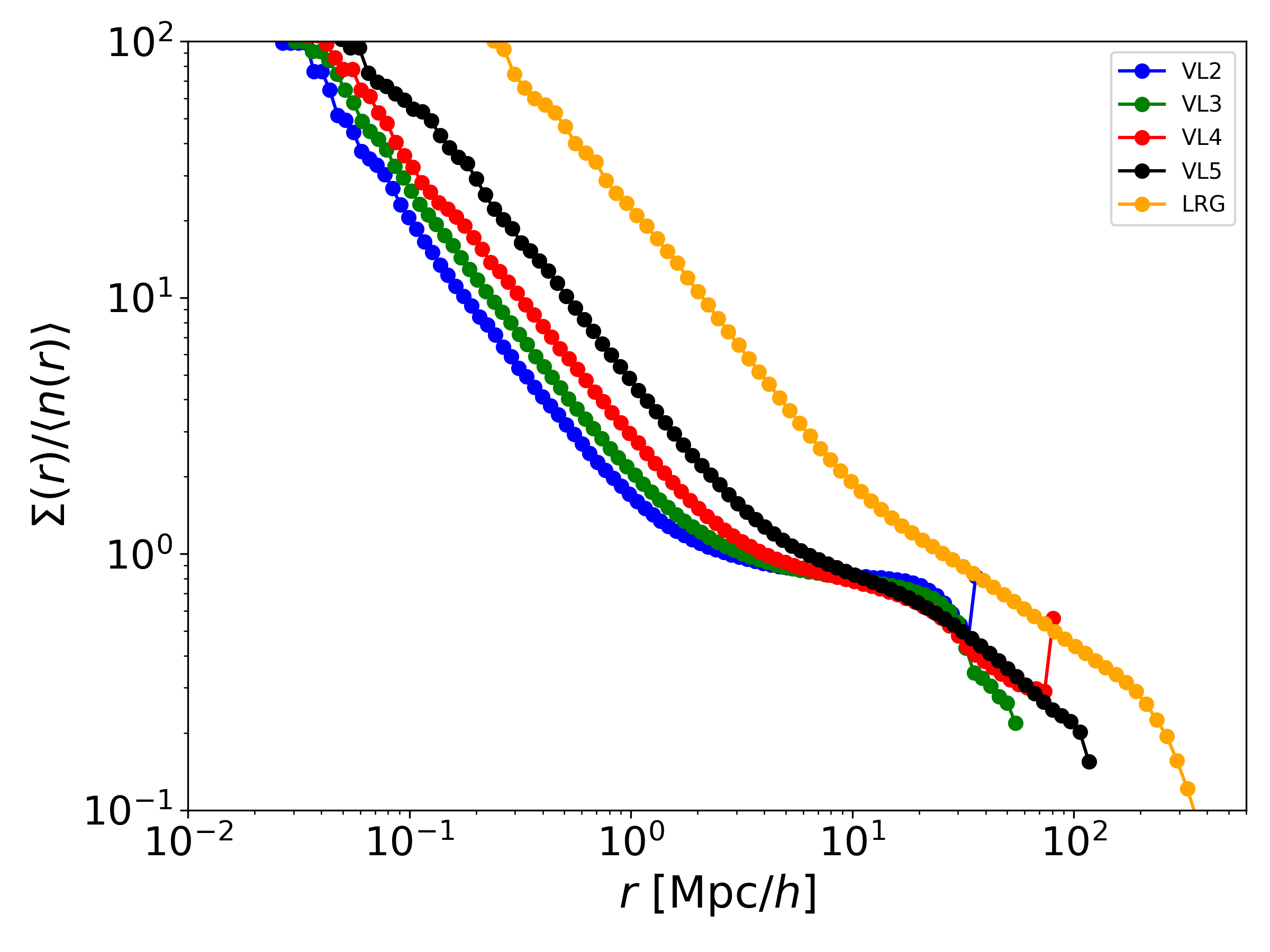}
\caption{Behavior of the relative standard deviation $\Sigma(r)/\langle n(r) \rangle$ as a function of the distance for the different samples. Up to scales $\sim 200$~Mpc$/h$ this quantity is of order one or larger, which signifies a lack of self averaging and the presence of large scale structures.}
\label{ave_stnd}
\end{figure}
\par

In the intermediate range of scales, the conditional density decays as $\langle n(r) \rangle \propto r^{-0.8}$; however, the decay tends to become progressively shallower as $r$ approaches the boundary of the sample, so that the deeper the sample, the larger the scale at which this deviation sets in. This trend is clearly visible in the logarithmic slopes shown in the bottom panels of Fig.~\ref{SL_bgs_1}. In all angular regions the onset of the slower decay shifts to progressively larger scales as the sample volume increases. We interpret this behavior as a manifestation of finite-size effects.

A way to minimize the finite--size effects associated with the overlap of sampling volumes at large radii---which we examine in more detail below---is to divide the sample into several disjoint angular regions. Specifically, for the LRGS we considered the region R25 and partitioned it into eight non-overlapping sub-regions by slicing its extent in right ascension. These sub-regions encompass approximately the same volume, since the maximum inscribed radius is constrained by the declination width, which is identical to that of the full R25 region.  

The results are shown in Fig.~\ref{SL_lrgs_2}. One may note that non-negligible fluctuations, at the level of $\sim 20\%$, characterize the variation of the average conditional density among these sub-regions. This finding is consistent with the behavior observed through the conditional variance (see Fig.~\ref{ave_stnd}), further indicating that large-scale fluctuations are not washed out by subdividing the sample.

The observation of a significant decay of the conditional density up to $r \sim 400$~Mpc$/h$ suggests that the galaxy distribution retains correlations on very large scales. For $r \gtrsim 200$~Mpc$/h$, however, the signal becomes increasingly affected by finite-size effects due to the limited sampling of the survey volume. Nevertheless, if the underlying fluctuations were truly small, such finite-size effects would be negligible. This indicates that the observed large-scale trends reflect genuine correlations, rather than mere sampling artifacts.

\subsection{The average conditional density in the Cold Dark Matter model}

In order to compare our findings with the predictions of a standard Cold Dark Matter (CDM) model, let us consider the simplified case of a power spectrum of the form
\begin{equation}
\label{pk}
P(k) = A \, \frac{k}{1 + \left( \frac{k}{k_t} \right)^{m}} \; ,
\end{equation}
where $A$ is a normalization constant, $k_t \approx 0.05\,h\,\mathrm{Mpc}^{-1}$ is the turnover scale, and $m \approx 2.5$ characterizes the small-scale (large-$k$) behavior.  
For clarity, we adopt a power spectrum without baryon acoustic oscillations, as these features are not essential to the main arguments of this work.  

The normalization constant $A$ is chosen such that $\overline{\xi}(r_0)=1$ for $r_0 = 9$ and $14$~Mpc$/h$, respectively, where $\xi(r)$ is the correlation function---the Fourier conjugate of $P(k)$---and
\begin{equation}
\overline{\xi}(r) = \frac{3}{r^{3}} \int_{0}^{r} \xi(y)\, y^{2}\,dy \; .
\end{equation}
The conditional average density, normalized by the (well-defined) mean density $n_0$ in this model, is then
\begin{equation}
\langle n(r) \rangle = n_0 \left[1 + \overline{\xi}(r)\right] \; .
\end{equation}

Note that Eq.~\ref{pk} corresponds to the prediction of linear theory. However, our interest lies in the behavior at scales $r > 50$~Mpc$/h$, whereas nonlinearities in standard CDM models emerge only at scales $r < 10$~Mpc$/h$, producing a modification of the small-scale slope of $\xi(r)$ without affecting the large-scale trends relevant here.

Results are shown in Fig.~\ref{SL_cdm}: one may note that there is a smooth and well-defined transition to homogeneity in both cases, occurring at $\sim 80$~Mpc$/h$ and corresponding to $\gamma \rightarrow 0$.  
The main difference with the data (see Fig.~\ref{SL_bgs_1}) appears for $r > 80$~Mpc$/h$, where the real samples exhibit both large fluctuations and a persistent positive exponent, $\gamma > 0$.

\begin{figure*} 
\includegraphics[width=0.5\textwidth]{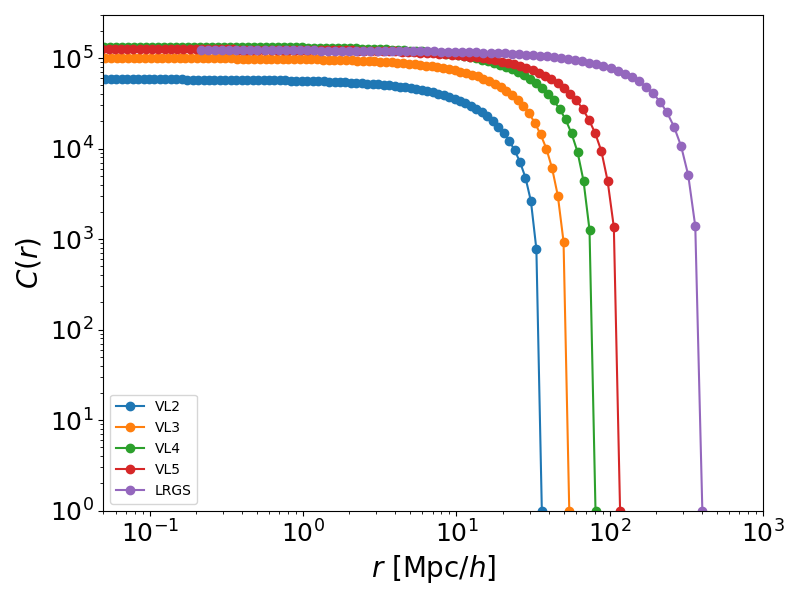}
\includegraphics[width=0.5\textwidth]{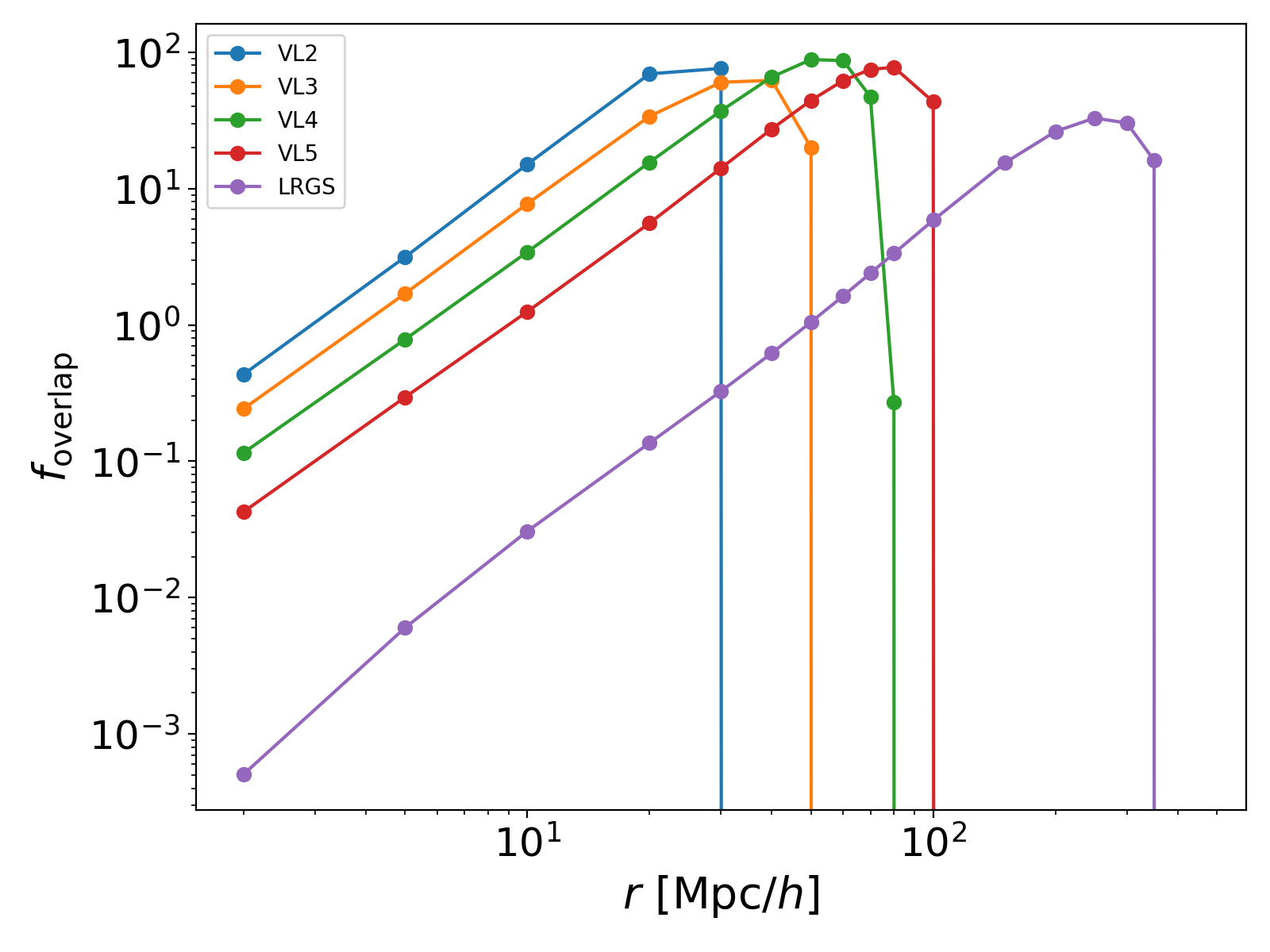}
\caption{
Left Panel: 
Behavior of the number of centers, $C(r)$, over which the conditional density $\langle n(r) \rangle$ is averaged
for the four samples VL2, VL3, VL4, VL5, LRGS in the same angular region R25. Right Panel: 
behavior  of $f_{\mathrm{overlap}}$ the total overlapping volume of all intersecting sphere pairs and the total volume of all spheres (see Eq.~\ref{eq:overlap}) for the same samples.
} 
\label{SL_bgs_2}
\end{figure*}

\subsection{Conditional variance}

The variance $\Sigma^2(r)$ measured in the R25 region for the four volume-limited subsamples extracted from the BGS and for the LRGS is shown in Figure~\ref{SL_bgs_5}. 
The difference in the small-scale slope of the variance  between the various samples arises from the different average distances between nearest-neighbor galaxies, denoted by $\Lambda$, which are in turn determined by the distinct luminosity selection criteria applied to each sample. For instance, in the BGS sample we find $\Lambda \approx 0.5$~Mpc/$h$, while in the LRGS sample $\Lambda \approx 20$~Mpc/$h$. The variance, similarly to the mean
(see Fig.\ref{SL_bgs_1}), exhibits a non-trivial power-law behavior on small scales,  
with $\Sigma^2(r) \propto r^{-3.5}$, and it departs from this regime on larger scales,  
where the scaling becomes $\Sigma^2(r) \propto r^{-2.4}$. 
The scale at which the slope changes depends on the sample, indicating that this feature is driven by finite–size effects.

At scales larger than the average distance between nearest neighbors, i.e., $r \gtrsim \Lambda$, where shot noise from sparse sampling becomes non-negligible, the variance exhibits a slower decay than expected for a homogeneous Gaussian field. Specifically, we find $\Sigma^2(r) \propto r^{-2.4}$, in contrast to the standard scaling $\Sigma^2(r) \propto r^{-3}$. This slower decay reflects the persistence of fluctuations over large distances, indicative of long-range correlations in the galaxy distribution. Such behavior is consistent with the evidence that clustering persists across a wide range of scales, without a clear transition to homogeneity.

Fig.~\ref{ave_stnd} shows the behavior of the ratio $\Sigma(r)/\langle n(r)\rangle$ for the different samples. 
The finite-size dependence of this quantity suggests the presence of non-trivial volume effects, likely driven by the inclusion of increasingly large structures as the sample volume increases. 
The consistency of this trend across both the BGS and LRGS samples further strengthens the evidence for a robust large-scale clustering signal. 
At large scales, this ratio reaches values of order $\sim 0.3$, indicating that relative fluctuations remain at the $\sim 30\%$ level even at the largest radii that are not clearly affected by finite-size effects (i.e.\ up to $\sim 200$~Mpc$/h$). 
Such a persistent level of variance demonstrates the absence of self-averaging --- reflecting the presence of large-scale structures within the sample --- and is fully consistent with the Gumbel-like form of the galaxy-count PDF, as we discuss below.

\subsection{Finite size effects}

Let us now explore in more detail the finite-size effects in the determination of the conditional density. The behavior of the number of centers, $C(r)$, over which the conditional density $\langle n(r) \rangle$ (see Eq.~\ref{eq:avecon}) is averaged, is shown in the left panel  Figure~\ref{SL_bgs_2} for the following cases the different samples in the angular region R25. 
The figure illustrates the typical trend in which $C(r)$ remains approximately constant at small and intermediate scales, but drops sharply as $r$ approaches the sample boundaries. This behavior reflects the progressive loss of fully contained spheres and highlights the increasing impact of edge effects at large separations.

As expected, deeper samples correspond to larger effective volumes and exhibit a smoother, less fluctuating behavior in the conditional density $\langle n(r) \rangle$. The scale beyond which $C(r)$ begins to decline marks the limit beyond which estimates of the conditional density become increasingly unreliable due to edge effects and reduced statistical sampling.

The right panel of Figure~\ref{SL_bgs_2} shows for the samples the behavior of the quantity denoted as $f_{\mathrm{overlap}}$, defined as the ratio between the total overlapping volume of all intersecting sphere pairs and the total volume of all spheres (see Eq.~\ref{eq:overlap}). As expected, the overlap increases rapidly when the radius $r$ of the spheres used to compute the conditional density approaches the radius of the largest sphere that can be fully contained within the sample volume.

The results presented in the panels of Figure~\ref{SL_bgs_2}  are both complementary and consistent: they indicate that measurements at scales approaching the size of the sample may be significantly affected by systematic effects. In particular, while statistical errors may underestimate the true uncertainties in this regime, a more robust approach is to assess the stability of the results by varying the sample size. In this context, the comparison of the conditional density and related quantities across the three selected angular regions provides a critical test of the robustness of the signal.

The behaviors of $C(r)$ and $f_{\mathrm{overlap}}(r)$ point to the same conclusion: as the boundary of the sample is approached, the volume average becomes increasingly poorly sampled, and the corresponding measurements are more prone to systematic biases. This is clearly demonstrated by comparing the results across the three angular regions: only by increasing the sampled volume does the signal become more stable and well-defined.

We emphasize that finite-size effects are particularly relevant in the presence of large-scale structures within the sample. In such cases, enlarging the volume does not merely reduce statistical noise, but is necessary to properly sample the underlying distribution. This is indeed the case for the present samples, where large-scale structures contribute non-trivially to the observed signal.

\begin{figure} 
\includegraphics[width=0.45\textwidth]{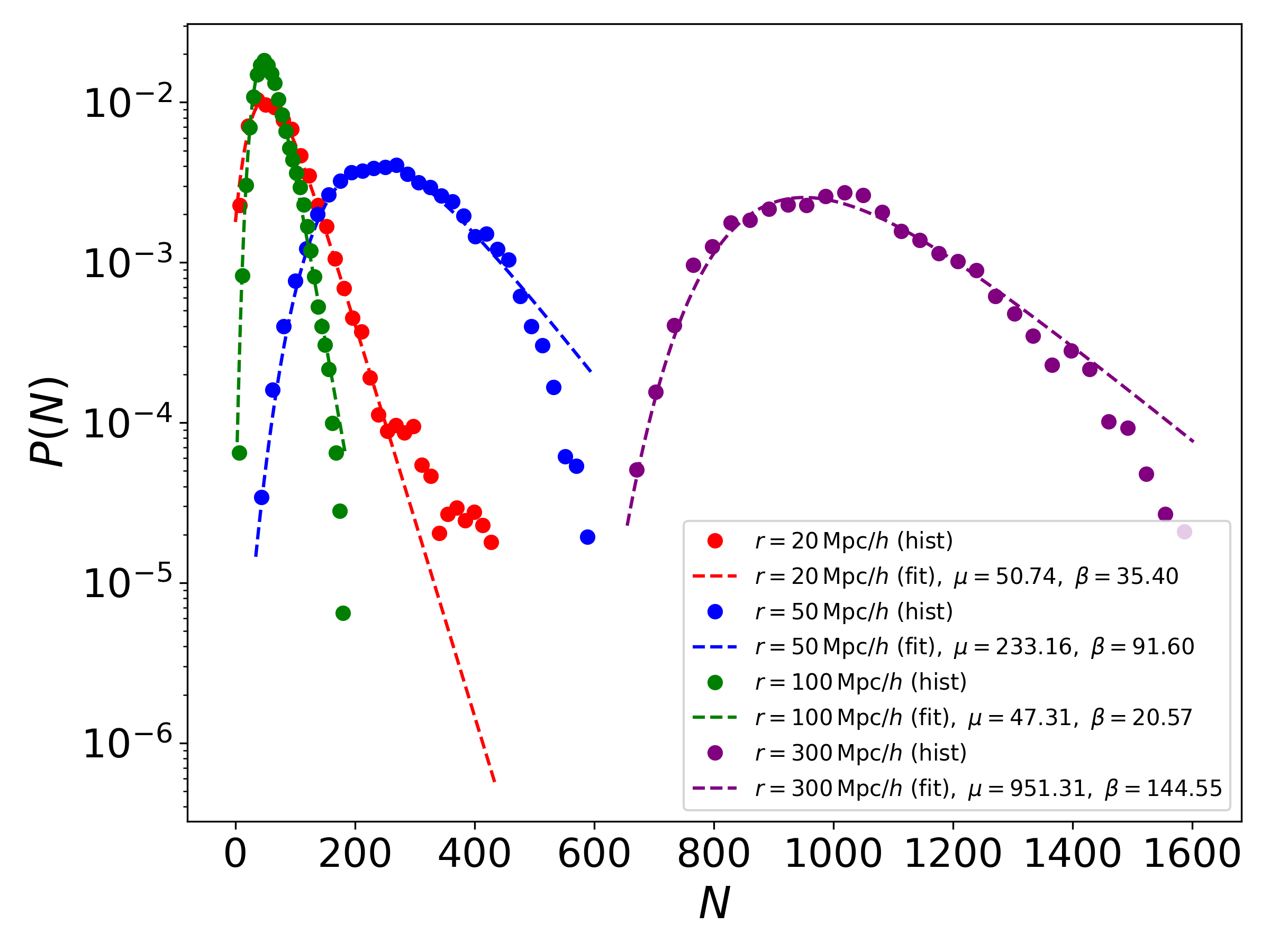}
\caption{
Probability density function (PDF) of the galaxy counts $P(N; r)$ in spheres 
of different radii and for different samples: 
top left: VL4 at $r = 20$~Mpc/$h$; 
top right: VL5 at $r = 50$~Mpc/$h$; 
bottom left: LRGS at $r = 100$~Mpc/$h$; 
bottom right: LRGS at $r = 300$~Mpc/$h$.  
The best-fit Gumbel distribution is also shown.
} 
\label{pdf_bgs_2}
\end{figure}

\subsection{Probability density function of fluctuations} 

Figure~\ref{pdf_bgs_2} shows the PDF $P(N,r)$,
where $N$ is the integrated number of points in a sphere of radius $r$
\[
N(r) = \int_0^r n(s) d^3s 
\]
for various values of the radius $r$ and for different samples. One may note that, in all cases, the Gumbel distribution provides an good fit to the data.

\begin{figure*} 
\includegraphics[width=0.5\textwidth]{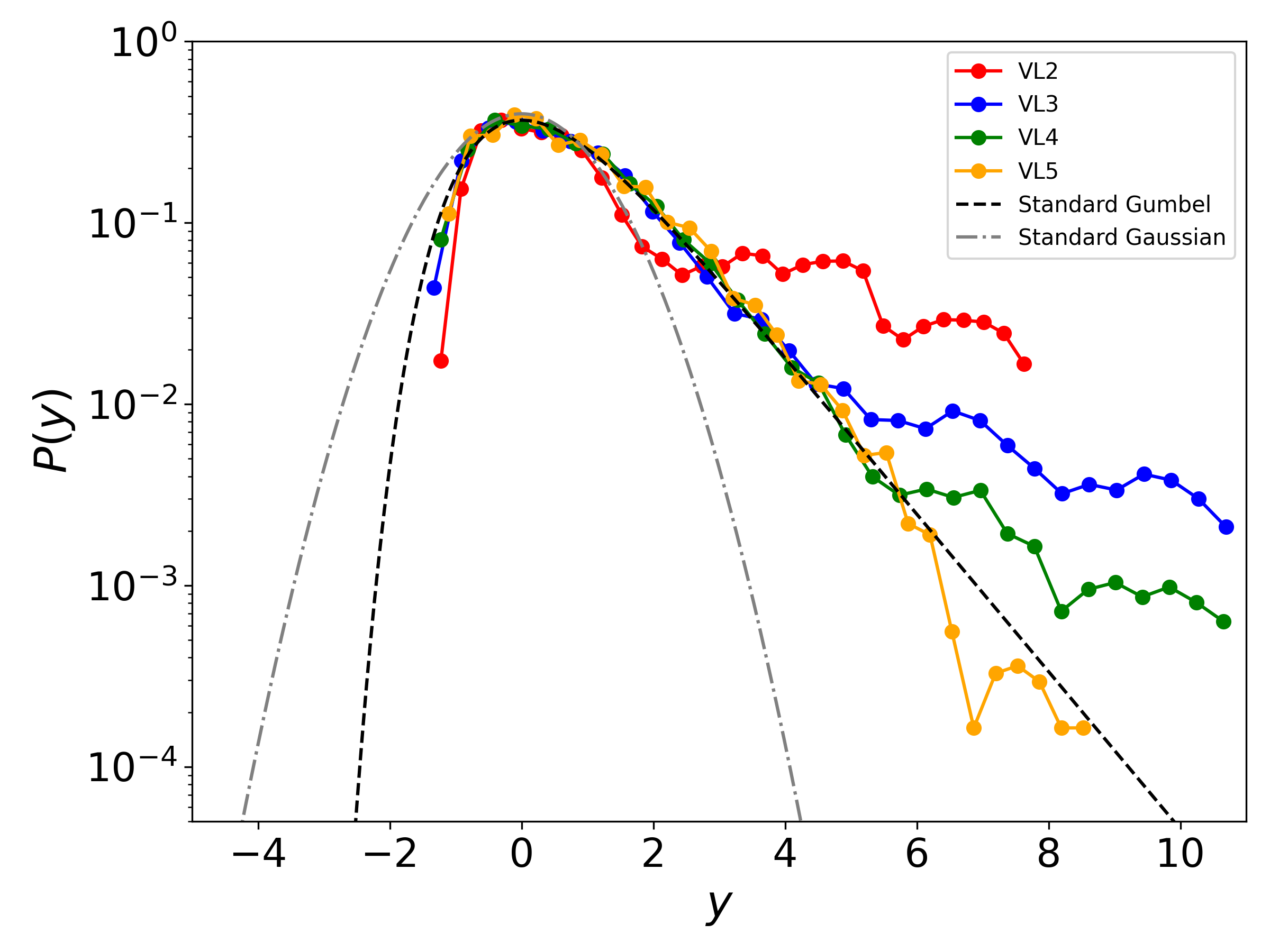}
\includegraphics[width=0.5\textwidth]{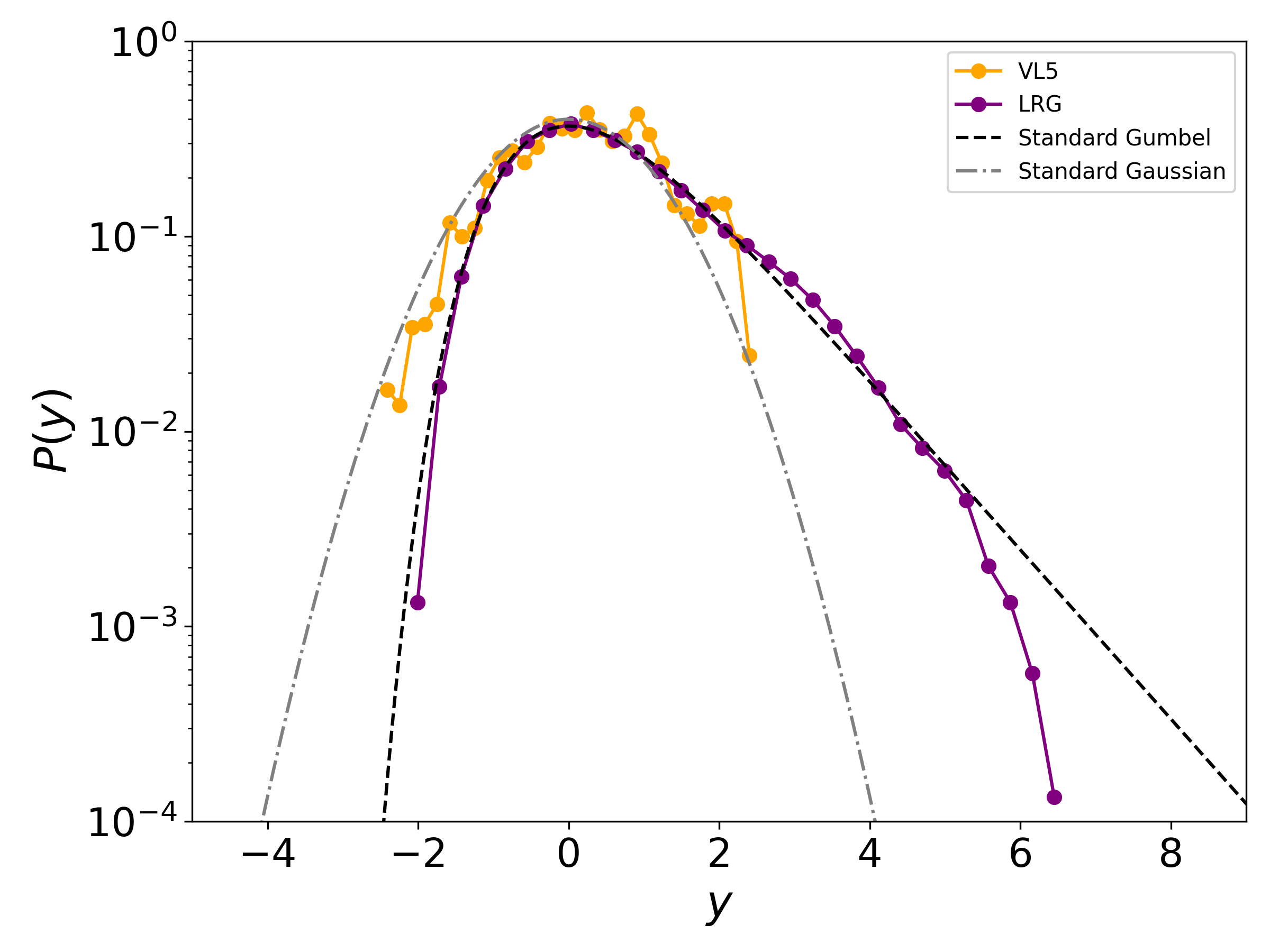}
\caption{Normalized PDF for $r = 20$ and $100$~Mpc/$h$ in different samples. The standard Gumbel and Gaussian functions are also reported as reference.} 
\label{pdf_lrg_2}
\end{figure*}

In order to compare different radii and samples we consider the 
PDFs  normalized according to 
\begin{equation}
\label{eq:norm} 
y = \frac{N - \mu}{\beta} \;,
\end{equation}
where $\mu$ and $\beta$ are the parameters obtained from the best fit to a Gumbel distribution.
Fig.~\ref{pdf_lrg_2} shows the results for $r = 20$ and $100$~Mpc/$h$ in different samples. One may note that, as the sample volume increases --- i.e.\ moving from VL2 to VL3, VL4, VL5, and finally to the LRGS --- the PDF is increasingly well approximated by the Gumbel distribution. A similar trend is observed when considering the different subsamples of the same dataset.  

One finds that, as the sample volume grows --- from VL2 to VL5 to the LRGS --- the PDF at a given radius becomes less affected by statistical fluctuations and develops a more pronounced tail. This behavior is consistent with the trends found for the first and second moments of the distribution, namely the conditional average density and its variance.  

At smaller radii (e.g., $r = 20$~Mpc/$h$), the measured PDF follows the Gumbel distribution more closely. This occurs because, at such scales, the averaging volume is smaller and more finely samples the local density variations, allowing a more accurate reconstruction of the underlying statistical fluctuations of $n$.

\begin{figure} 
\includegraphics[width=0.45\textwidth]{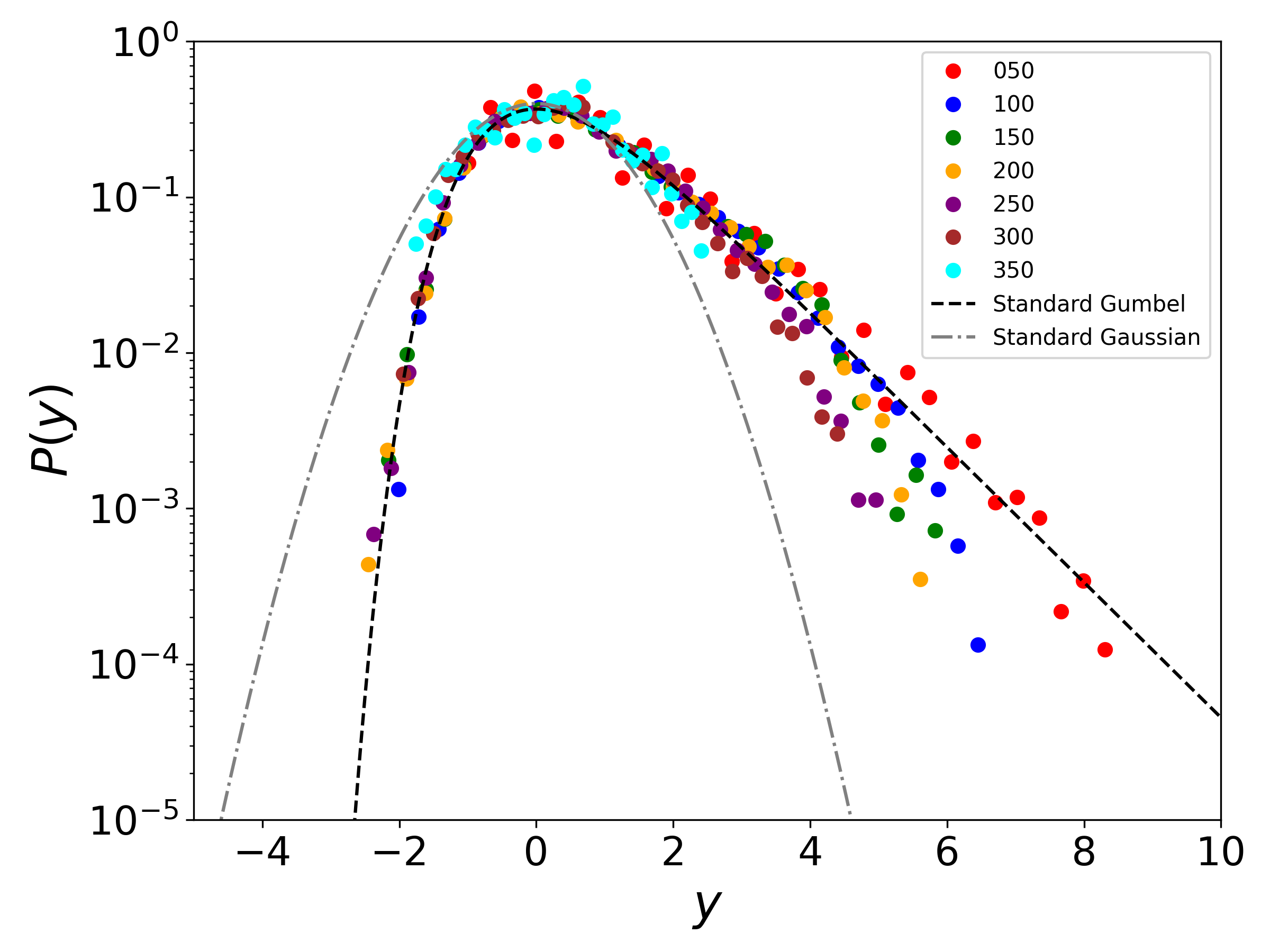}
\caption{ Collapse plot of the normalized  PDF for the LRGS  in region R25, for different values of the sphere radius $r$.
The standard Gumbel and Gaussian functions are also reported as reference. 
} 
\label{pdf_lrg}
\end{figure}

Figure~\ref{pdf_lrg} shows the collapse plot of the normalized  PDF  for the LRGS in region R25, for different values of the sphere radius $R$. One may note a trend similar to that observed in the BGS samples: the larger the sampling volume, the narrower the range of the normalized variable $y$ spanned by the measured histogram. This behavior is a typical finite-size effect, reflecting the fact that the volume average is systematically influenced by the limited extent of the sample. As the sample volume increases, fluctuations are better sampled, and the PDF converges more closely to the underlying statistical distribution.

\section{Conclusions}
\label{sec:concl}

In summary, the behaviors of the conditional density, the variance, and the PDF of fluctuations can all be interpreted as manifestations of the non-homogeneity of the galaxy distribution. These results are in agreement with those detected in smaller samples by \cite{Antal_etal_2009} (and references therein).

Whether the observed power-law exponent $\gamma \approx 0.8$ continues to characterize the conditional density at scales $r \gtrsim 200$~Mpc$/h$ remains uncertain. A slow transition to homogeneity, or a weaker dependence on scale—possibly logarithmic—cannot be excluded. Nonetheless, our results clearly demonstrate that a transition to spatial homogeneity has not yet occurred within the explored range, up to $\sim 400$~Mpc$/h$. This conclusion contrasts with those of other authors --- for example, \citet{Hogg_etal_2005}, who also studied the conditional density while controlling for luminosity selection effects and boundary conditions. The deeper DESI samples allow for a more accurate characterization of finite-size effects, which likely influenced earlier determinations and may have led to an underestimation of the scale of inhomogeneities.

The persistence of correlations and the absence of a clear crossover toward homogeneity --- i.e.\ toward $\gamma = 0$ --- at scales of several hundred megaparsecs challenge the standard assumption of large-scale homogeneity. A Gaussian random field, as typically assumed in standard cosmological models on large scales, would predict a constant conditional density for radii larger than $\sim 20$~Mpc$/h$ \citep{SylosLabini_2011}, a more rapid decay of the variance as $\propto r^{-3}$ (and a normalized variance decaying as $\propto r^{-4}$ for $r \gtrsim 100$~Mpc$/h$; see \citealt{glasslike}), and a Gaussian form for the PDF of counts-in-spheres.

These findings pose a fundamental challenge to standard cosmological models, as the FRW  metric itself is derived under the assumption of spatial homogeneity --- an assumption not supported by the large-scale galaxy distribution observed in the DESI sample.

One may argue for the existence of a uniform, unobservable background that fills space, thereby justifying the assumption of a well-defined average mass density. However, even under this hypothesis, it remains extremely challenging to reconcile the observed spatial correlations --- and hence the presence of large-scale structures extending over hundreds of megaparsecs --- with the standard paradigm of galaxy formation. Within this framework, gravitational dynamics in an expanding universe can generate nonlinear structures only up to scales of approximately $\lambda_0 \approx 10-20$~Mpc$/h$; beyond this scale, clustering remains in the linear regime, and the density field is expected to retain the same correlation properties as the initial conditions, i.e. anti-correlations on scales $r \gtrsim 100$~Mpc$/h$.

Therefore, the results presented here raise fundamental questions about the validity of the standard cosmological scenario, particularly concerning the assumption of large-scale homogeneity in the matter distribution of the universe. Forthcoming data releases from DESI and other ongoing galaxy surveys~\citep{Euclid_Collaboration_2022} will provide access to even larger spatial volumes. This will enable us to test the robustness of the present findings with significantly larger samples and improved statistical analyses, potentially corroborating the presence of persistent large-scale correlations in the galaxy distribution.
\bigskip

\section*{Data Availability}
{
The DESI data employed in this work are publicly available through the DESI collaboration. 
The analysis pipeline developed for this study is available from the authors upon request.
}
\bigskip

\begin{acknowledgements}

This work is dedicated to the memory of Yurij V. Baryshev. 

F.S.L. is grateful to Michalel Joyce for a list of comments and suggestions and to Giordano De Marzo and Andrea Gabrielli for many insightful and stimulating discussions.

This research used data obtained with the Dark Energy Spectroscopic Instrument (DESI). DESI construction and operations is managed by the Lawrence Berkeley National Laboratory. This material is based upon work supported by the U.S. Department of Energy, Office of Science, Office of High-Energy Physics, under Contract No. DE--C02--05CH11231, and by the National Energy Research Scientific Computing Center, a DOE Office of Science User Facility under the same contract. Additional support for DESI was provided by the U.S. National Science Foundation (NSF), Division of Astronomical Sciences under Contract No. AST-0950945 to the NSF's National Optical-Infrared Astronomy Research Laboratory; the Science and Technology Facilities Council of the United Kingdom; the Gordon and Betty Moore Foundation; the Heising-Simons Foundation; the French Alternative Energies and Atomic Energy Commission (CEA); the National Council of Humanities, Science and Technology of Mexico (CONAHCYT); the Ministry of Science and Innovation of Spain (MICINN), and by the DESI Member Institutions: www.desi.lbl.gov/collaborating-institutions. The DESI collaboration is honored to be permitted to conduct scientific research on I'oligam Du'ag (Kitt Peak), a mountain with particular significance to the Tohono O’odham Nation. Any opinions, findings, and conclusions or recommendations expressed in this material are those of the author(s) and do not necessarily reflect the views of the U.S. National Science Foundation, the U.S. Department of Energy, or any of the listed funding agencies.
\end{acknowledgements}

 \bibliographystyle{aa}

\end{document}